\documentclass[12pt,preprint]{aastex}

\usepackage{lscape}

\def\spose#1{\hbox to 0pt{#1\hss}}
\def\lta{\mathrel{\spose{\lower 3pt\hbox{$\mathchar"218$}}
     \raise 2.0pt\hbox{$\mathchar"13C$}}}
\def\gta{\mathrel{\spose{\lower 3pt\hbox{$\mathchar"218$}}
     \raise 2.0pt\hbox{$\mathchar"13E$}}}
\def\p0{\phantom{0}}
\def\pd0{\phantom{00}}
\shorttitle{Planet Hosting Multiple Star Systems}
\shortauthors{Cuntz et al.}
\graphicspath{{./}{figures/}}

\begin{document}

\title{
An Early Catalog of Planet Hosting Multiple Star Systems \\
of Order Three and Higher
}

\author{M. Cuntz, G. E. Luke, M. J. Millard, L. Boyle, S. D. Patel}


\affil{Department of Physics, Box 19059}
\affil{University of Texas at Arlington, Arlington, TX 76019;}
\email{cuntz@uta.edu; gregory.luke@mavs.uta.edu; matthew.millard@mavs.uta.edu; lindsey.boyle@mavs.uta.edu; sdp3361@mavs.uta.edu}











\begin{abstract}
We present a catalog (status July 1, 2022) of triple and higher order systems
identified containing exoplanets based on data from the literature, including
various analyses.  We explore statistical properties of the systems with
focus on both the stars and the planets.  So far, about 30 triple systems and
one to three quadruple systems, including (mildly) controversial cases,
have been found.  The total number of planets is close to 40.  All planet-hosting
triple star systems are highly hierarchic, consisting of a quasi-binary
complemented by a distant stellar component, which is in orbit about the common center
of mass.  Furthermore, the quadruple systems are in fact pairs of close binaries
(``double-doubles"), with one binary harboring a planet.
For the different types of star-planet systems, we introduce a template for the
classifications of planetary orbital configurations in correspondence to the hierarchy
of the system and the planetary host.  The data show that almost all stars are
main-sequence stars, as expected.  However, the stellar primaries tend to be more massive
(i.e., corresponding to spectral types A, F, and G) than expected from single star
statistics, a finding also valid for stellar secondaries but less pronounced.
Tertiary stellar components are almost exclusively low-mass stars of spectral type M.
Almost all planets have been discovered based on either the Radial Velocity or the
Transit method.  Both gas giants (the dominant type) and terrestrial planets
(including super-Earths) have been identified.  We anticipate the expansion
of this data base in the light of future planetary search missions.
\end{abstract}





\section{Introduction}


Comprehensive studies of star-planet systems have become a main segment of
contemporaneous astronomy, astrophysics, and astrobiology.  Since the first
detection of a planet beyond the Solar System in orbit about a solar-type star
\citep{mayo95}, the number of confirmed exoplanets\footnote{For updated
information see, e.g., {\tt http://exoplanet.eu} and {\tt http://exoplanets.org}}
is currently exceeding 5000.  Note that most
planets are hosted by single stars, although there is a notable
number of planets (which is about 100, in part depending on the cutoff regarding
the stellar separation distance) that are members of stellar binary systems
\citep[e.g.,][]{duqu91,pati02,egge04,lada06,ragh06,ragh10,roel12,wang15a,wang15b,pila19}.
However, the number of planets found to be hosted by higher order systems
is relatively small, i.e., about 40 for triple and quadruple systems
combined, with the exact number depending on whether some controversial
or unconfirmed cases are included.  These kinds of systems are the focus
of the present study.

Note that a considerable fraction of stellar systems is composed
of multiple stars, which have previously been the topic of intense research,
encompassing stellar structure analyses, orbital stability studies, and research
devoted to evolution (see below).   Following \cite{toko14a,toko14b},
single stars account for approximately half, binaries a third, and triples
less than a tenth.  Comparable statistics, including studies on the origin
of triple and quadruple stellar systems, have been given by \cite{toko08} and
\cite{eggl08}.  Previous studies about the stability of planets in triple and
higher order stellar systems have been given by \cite{ford00}, \cite{mard01},
\cite{verr07}, \cite{hame15}, \cite{corr16}, \cite{buse18}, and \cite{myll18}.

Regarding planets in binary systems, two different kinds of cases have been identified
\citep{dvor82}.  First, planets may orbit one of the binary components; those are categorized
as planets in an S-type orbit (see Fig.~1).  In those systems the other stellar component is at a notable
distance; however, it may act as a perturbator.  Second, planets may orbit both binary
components; those are said to be in P-type orbits.  This type of nomenclature has also
been adopted by, e.g., \cite{buse18} for triple stellar systems; it will be applied and
augmented in this study. Previously, hierarchical triple star systems have also been
studied by \cite{verr07} with a focus on single planets in orbit about inner binaries.
For studies of possible habitability for planets in S-type and P-type orbits see, e.g.,
\cite{cunt14}, and references therein.

\newpage

\cite{toko08} conveyed comparative statistics and comments on the origin of triple
and quadruple stellar systems.  For triple and quadruple systems, they found notably
different statistics for the orbital periods and mass ratios.  Regarding the quadruple
systems, they identified a relative high abundance of $\epsilon$~Lyrae-type systems
(double binaries, also called ``double-doubles") with similar masses and inner periods.
However, the outer and inner mass ratios in triple and quadruple stars were not
mutually correlated.  They also argue that the origin of triple and quadruple
star systems could perhaps be explained via rotationally driven (cascade)
fragmentation possibly followed by migration of inner and/or outer orbits
to shorter periods.

Besides observational verifications indicating that planets are able to exist
in binaries and multiple stellar systems, additional support about the abundance
and significance of those planets stems from the discovery of protoplanetary
disks in those systems.  Besides the plethora of results on binaries, see, e.g.,
\cite{rodr98}, \cite{jens14}, \cite{czek19}, and references therein --- encompassing
different aspects such as planet formation, disk alignment, and evolution ---
a rare example of a disk in a triple star system was recently identified as well.
\cite{bi20} found that GW~Ori, a hierarchical triple system, possesses a rare
circumtriple disk.  The authors noted three dust rings in the GW~Ori disk at ${\sim}46$,
${\sim}188$, and ${\sim}338$~au, with an estimated dust mass of 74, 168, and 245 Earth masses,
respectively.  The data also indicate complex disk dynamics initiated by the
various stellar components.  Subsequent work by \cite{smal21} conveyed evidence
of disk breaking that according to the authors is likely caused by undetected
planets, which if confirmed would constitute the first planet(s) identified in
a circumtriple orbit.

The present-day record regarding stellar multiplicities are
the currently known two septuple systems AR Cassiopeiae and $\nu$~Scorpii, see
\cite{toko97} and \cite{eggl08}, including references therein.  Both systems are
highly hierarchic, with both containing different types of stars (although the
properties of many components still await classification), including
main-sequence stars.  Higher multiplicities may in fact be possible; in fact,
there may be a smooth transition to (though perhaps loosely) gravitationally bound
``open clusters".  Owing to the complex patterns of gravitation interaction in
those systems, possible exoplanets hosted in high-order systems may be restricted
to (probably tight) S-type configurations and to P-type configuration with respect
to quasi-binaries.  However, no detections have been made.

More recently, work pertaining to planet-hosting star systems has been given by, e.g.,
\cite{mugr19} and \cite{lest21}.  \cite{mugr19} presented a new survey that explores
the second data release of the ESA {\it Gaia} mission, in order to search for stellar companions
of exoplanet host stars, located at distances closer than about 500~pc around the Sun.
In total, 176 binaries, 27 hierarchical triples, and one hierarchical quadruple system have
been detected among more than 1300 exoplanet host stars.  As part of their study, they
examined dynamical aspects of the systems as well as the associated mass distribution.
\cite{lest21} reported speckle observations of TESS exoplanet host stars, with a focus
on stellar companions at 1--1000~au, including implications for small planet detection.

The purpose of this work is to provide a repository of data for triple and quadruple
star-planet systems as known to date, together with elements of analysis and interpretation.
In Section 2, we report on the data acquisition for the various star-planet systems.
We also offer a template of system classification, which particularly considers the
various types of planetary orbits.  Additionally, we discuss cases of controversies
and planet retractions as reported in the literature.  In Section 3, we provide
selected statistical analyses and interpretation, encompassing both the planets
and the host stars.  As a case study, detailed information about the Alpha \& Proxima
Centauri system, including aspects of orbital stability and habitability, is presented
in Section 4.  Additional analyses are given in Section 5, whereas
Section 6 conveys our summary and conclusions.
See Appendix A for the list of acronyms.


\section{Data Acquisition}

\subsection{Approach and Attributes}










The aim of this study is to present a catalog of planet-hosting multiple star systems
identified to contain exoplanets (see Table 1).  The underlying data have been obtained from the literature.
They include data on the planets and the stellar components, especially the planetary host stars,
as well as information on the stellar distances and the composition of the stellar systems.
For star-planet systems that are outside of the scientific community's main focus, thus lacking
notable updates, the original data provided at the time of discovery have been used.  In other cases,
as, e.g., the Alpha Centauri system, updates became available when additional system planets
have been discovered, often leading to the refinement of the data of previously detected planets.

Figure 2 and 3 depict the distribution of currently known planets in triple and quadruple stellar
systems in the two celestial hemispheres.  Certain (the overwhelming majority) and uncertain detections
are identified as such.  The images were produced using the StarCharter software\footnote{Source of data:
{\tt https://github.com/dcf21/star-charter}}.  Each image is a chart of all of the stars, above a
threshold magnitude, and constellations from its database.  Each image also plots the path of the
celestial equator, ecliptic plane, and the galactic plane.  Planetary detections, including the small
number of unconfirmed cases, at present occur in 22 out of the 88 constellations.  Unsurprisingly,
a high accumulation of systems takes place at the Summer Triangle, a consequence of the
multi-year contributions by the {\it Kepler} mission.  Information about the system
coordinates\footnote{Source of data: {\tt https://heasarc.gsfc.nasa.gov/cgi-bin/Tools/convcoord/convcoord.pl}}
is given in Table~2.

Figure~4 conveys information on the year of discovery\footnote{The year of discovery is set as the year
of publication concerning the respective planet.} for planets hosted by triple or quadruple star systems.
The first confirmed discovery has been 16~Cygni~Bb \citep{coch97}, a planet in a highly elliptical 
orbit hosted by a solar-type star being part of a triple star system.  In total, there are 27 confirmed
triple stellar systems with planets and one confirmed planet-hosting quadruple stellar system,
which is Kepler-64 \citep{schw13}, amounting to 39 confirmed planets in multiple stellar systems.
In fact, there are 7 multi-planetary systems; the total number of planets in those systems is 18.  Both
the confirmed and nonconfirmed planet-hosting quadruple stellar systems are so-called ``double-doubles",
i.e., pairs of stellar binaries, with one of them hosting a planet.  Regarding the triple star systems,
the current record indicates about one discovery per year over a timespan of 25 years, with
a notable increase within the last five years.  Actually, two discoveries already occurred in the
first few months of 2022, which are Proxima~Centauri~d and LTT~1445Ac; see \cite{fari22}
and \cite{wint22} for details.

Furthermore, we provide a list of the distances to each planet-hosting star system
(see Table~1 and Fig.~5).  The distance data mostly originate from the NASA Exoplanet Archive.
This archive relies on data from the TESS Input Catalog (TIC), Version 8.  The TIC uses
parallax measurements from {\it Gaia} to estimate the distance to the star systems; see
\cite{stas19}.  However, there are a few systems, i.e., HD 188753, KOI-5, and KIC 7177553,
where no information could be located in the archive.  Thus, for HD 188753 and KOI-5, the
distance reference of \cite{bail18} was used by utilizing their VizieR Catalogue\footnote{Source of data:
{\tt http://vizier.u-strasbg.fr/viz-bin/VizieR?-source=I/347}}.  KIC 7177553 appears to be
unavailable in both the NASA Exoplanet Archive and the VizieR Catalogue.  In this case, we quote the distance
value given by \cite{lehm16}.

The closest planet-hosting multiple star system is Alpha \& Proxima Centauri at a distance of 1.302~pc
{\citep{brow18}}\footnote{{\it Gaia} Data Release 2; see {\tt http://gea.esac.esa.int/archive/documentation/GDR2/index.html}}
--- a finding that will continue to persist considering that this system is closest to Sun-Earth.  However,
various planet-hosting triple star systems (certain detections), which are WASP-12, Kepler-13, and KOI-5,
were also discovered at relatively large distances; they are located at distances of 427~pc, 519~pc, and 547~pc,
respectively (see Table~1); for KOI-5, the existence of the planet was suggested by \cite{hirs17} and confirmed
by \cite{ciar21}.  A statistical analysis indicates that the mean and median values for the distances of
these kinds of systems are given as 45~pc and 120~pc, respectively.  Note that the significant skewness
of that distribution is due to a small number of large-distance outliers associated with successful detections
by SuperWASP and {\it Kepler}.

Table~3 and Fig.~6 give information on the method of discovery for the various exoplanets,
as well as the primary astronomical facility.  Regarding the latter, major contributions
have been made by La Silla, the Keck, Lick, and McDonald Observatories, as well as the {\it Kepler}
mission.  Most recently, contributions were also made by K2 and TESS.   Concerning
planetary discoveries, the dominant method has been the Radial Velocity method, prevalent especially
during the early phases of the planet discovery history and thus yielding 21 confirmed planets
in triple systems.  The second most important method has been the Transit method, also utilized by
{\it Kepler}, with an output of 16 confirmed planets in triple systems.  Minor contributions
have been made through direct imaging and by astrometry and eclipse timing.

Both gas giants (the dominant type) and terrestrial planets (including super-Earths) have been
identified (see Fig.~7).  The Radial Velocity method led to the discovery of many Jupiter-type
planets, noting that with respect to triple star systems, 24 Jupiter-type planets have been found.
However, in the absence of alternate methods, only minimum values for the masses could be
ascertained.  Earth-mass and sub-Earth-mass planets have been detected in the systems of Alpha Centauri
(see Sect.~4), Kepler-444, and LTT 1445; see \cite{fari22}, \cite{camp15}, and \cite{wint19} for
details and results.


\subsection{System Templates and Classifications}

\subsubsection{General Remarks}





This work is aimed at presenting and evaluating planet-hosting triple and
quadruple stellar systems.  A careful review of the data and literature shows
that most of those systems are genuine, which means that both the proposed
planet(s) and the respective stellar components can be assumed as confirmed;
see Table~4 and 5.  In those cases, there is no sincere controversy
on the existence of the various system components (OK case).  However, there
are some systems where this is not the case, referred to as Case~1, 2, and 3,
respectively.  In that regard, the following nomenclature is used.

Case~1 means that all stars have been confirmed and at least one exoplanet
has been confirmed; however, there is at least one other unconfirmed system planet
or stellar component.  Case~2 means that there is an unconfirmed stellar component
besides (at least) one confirmed  exoplanet.  Thus, the system may instead be a
planet-hosting binary (if a suggested triple star system) or a planet-hosting
triple star system (if a suggested quadruple star system).  Finally,  Case~3
means that all stellar components are confirmed; however, the proposed exoplanet
is still unconfirmed.  This kind of nomenclature is consistently used through
this work, including the various statistical analyses.

It is also found that all triple star systems as considered are hierarchic
in nature.  In those systems, like any other kinds of systems,
each star orbits the system's center of mass.  However, in a hierarchic triple
star system (see Fig.~1), two of the stars form a close binary, and the third orbits
this pair at a distance much larger than the distance of binary separation;
see, e.g., \cite{eggl95} and subsequent work for the examinations of triple system
stability and related topics.  Table~6 provides information on the configuration
of the triple stellar systems as discussed in this study as well as on the stellar spectral
types.  References of the latter are typical included in the publication of the planetary
discovery (see Table 1) or available through
SIMBAD\footnote{See {\tt http://simbad.u-strasbg.fr/simbad/}}.
We also give information on
small and large separation distances of the stellar components, given as 
${a_{\rm bin}}^{(1)}$ and ${a_{\rm bin}}^{(2)}$ as obtained from the literature
(see Sect. 2.2.3).

\subsubsection{Comments on Planetary Orbits}

When attempting to identify and analyze systems\footnote{
Stellar components are denoted by using the suffixes A, B, C (etc.) and system planets
are denoted by using the suffixes b, c, d (etc.).  More intricate notations are used
for, e.g., circumbinary planets.  Setting adequate standards of notation is the
charge of the IAU Commission G1, which is striving for optimal
consistency, a challenging task particularly for high-order systems.
Relevant publications in this context include \cite{hess10} and \cite{toko14a,toko14b}.}
containing an abundant number of
stellar and/or planetary components, especially those identified by different groups
or individuals, their identity can be difficult to distinguish.  For instance,
an AC-B system\footnote{Systems denoted as AB-C or AC-B represent, in principle,
the same kind of configuration, which is a close binary given as AB or AC, respectively,
and a distant component referred to as component C or B, respectively.
The different usage of names is usually due to the observational history of those stars.} ---
assumed as hierarchic --- is named that way
because the C component has been found to orbit the component A (the primary) but
this discovery took place after the confirmed existence of the distant component B.
Thus, the C component would constitute the second stellar component in that system
irrespectively of the stars’ discovery history.  Regarding our study, we have adopted
this terminology focusing on the order of the stellar components rather than the name
given to them.  Consequently, for the case of planetary orbits in an AC-B system,
if a planet orbited A, it would be S1, if it orbited C, it would be S2, and if it orbited B,
it would be S3.  Equivalent terminology applies to any variation of the traditional
AB-C or A-BC system.  Here, a planet in orbit about the B and C component would be
referred to as S2 and S3, respectively.

A planet in a quasi-circumbinary orbit about the components A and C in an AC-B system
is referred to as P12.  However, due to the lack of orbital stability, P13 and P23
planets in an AC-B system, in orbit about AB and CB, respectively, would be impossible
in this kind of hierarchic triple system.  In case of non-hierarchic triple systems, the
principal possibility of P13 and P23 planets may exist, although detailed stability
analyses would be required to verify their existence.  Planets orbiting all three
stellar components at once, expected to require wide orbits, may occur in
non-hierarchical and mildly hierarchical triple stellar systems; they would be
referred as T123 planets (``circumtrinary" planets).  However, those cases still
await observational verification.

Information regarding the triple star systems considered in this study is given in
Table~7.  It is found that for both confirmed and unconfirmed systems the overwhelming
case is that of S1 with 28 and 4 identifications, respectively; see Table~7.  For
confirmed cases, the second most common case is S3 followed by S2, with 7 and 3
identifications, respectively.  For confirmed systems, no P12 case is found; however,
there is one unconfirmed P12 case, which is HW~Virginis.

The terminology used in this work bears some similarities to that previously adopted
by \cite{buse18}, although some notable differences exist. They
use very similar definitions for their planet orbits; however, their paper focuses
on studying one kind of configuration for triple systems, which is a hierarchic in
a particular manner. Thus, the inner binary is orbited by a third component outside
of the quasi-binary. Here we adopt a different kind of P-type orbital definition
as we account for both inner and outer binary pairs as seen in a possible AB-C
or A-BC system. Moreover, \cite{buse18} do not distinguish if they are
tracking all planets located within a system or simply noting the orbital type
of a confirmed planet in the system. In contrast, our study focuses on
a more comprehensive picture of possible triple stellar system configurations
instead of a single setup.

\subsubsection{Validation of the Stellar Components}

A crucial aspect of this study is to corroborate the existence of the
previously identified stellar system components.  Hence, we sought
identifying consensus in the scientific community, but also paid
heightened attention to controversial cases.  One of those cases
appears to be HD 4113 where there are open questions about one
of the system components (see Sect. 2.3).  In addition, there had
been a retraction pertaining to the HD~131399 system (see Sect. 2.4.2).

Naturally, systems where stellar components might be in jeopardy of
not being fully gravitationally bound to the system are those with
the largest values for ${a_{\rm bin}}^{(2)}$, the larger of the two
separation distances (semimajor axes); see Table 6.  Currently,
there are four cases of ${a_{\rm bin}}^{(2)} \gta 3000$~au, which are
HD~126614, HD~40979, HD~196050, and the Alpha \& Proxima Centauri
system.   For the latter, ${a_{\rm bin}}^{(2)}$ is identified as
approximately 8200~au based on recent ALMA data \citep{akes21}.

HD~126614 is an established triple star system whose outer component,
also known as LP 680-57, was first reported in 1960 with the W. J. Luyton
proper motion catalog \cite{maso01}; its angular separation is given as
$41.90^{\prime\prime}$ from the primary.  Work by \cite{lodi14} confirmed
that the system is gravitationally bound.  Regarding HD~40979, the
composition and connectedness have been examined by \cite{mugr07} who
concluded that the outer component, despite an angular separation of
$192.5^{\prime\prime}$ from the primary, is part of that system.
Moreover, HD~196050 has been identified as a gravitationally
bound triple star system as well.  In this case, the angular
separation of the outer component is given as $10.80^{\prime\prime}$,
corresponding to a projected separation of $7511 \pm 22$~au \citep{mugr05}.

Detailed comments about the Alpha \& Proxima Centauri system have been
given in Sect.~4.  In 2017 Radial Velocity measurements have been
precise enough to demonstrate that Proxima Centauri is gravitationally bound to
Cen~AB.  Proxima Cen is currently about 13,000~au away from the Cen~AB
owing to the relatively high orbital eccentricity given as
$0.497_{-0.060}^{+0.057}$ \citep{akes21}.  Due to their closeness to Earth
and the orbital eccentricities of Cen~AB and Cen~C, the position angles
continuously change throughout their projected orbits, whereas the orbital
parameters remain largely unaltered.  Recently, the system's
orbital stability has been reconfirmed by \cite{boyl21}.

In Table~6, we list ${a_{\rm bin}}^{(1)}$ and ${a_{\rm bin}}^{(2)}$
for the various systems, the small and large separation distances
of the stellar components, respectively, as obtained from the literature.
They are often relatively uncertain as they depend on the stellar
proper motions, among other quantities; in fact, they are typically also be
affected by projection effects.  For example, \cite{colt21} assessed
the existence of bound stellar components to {\it Kepler} exoplanet host stars
based on Speckle Imaging allowing insights into the nature of many of those
systems.  We also assessed the ratios $q_{\rm bin} = 
{a_{\rm bin}}^{(2)} / {a_{\rm bin}}^{(1)}$ for the various planet-hosting
triple star systems.  In that regard, at present, no compact systems with 
$q_{\rm bin} \lta 3$ have been identified, which also affects the types
of permissible planetary orbits.



\subsection{Controversies}

For some of the triple and quadruple systems, as listed in Table 5,
ongoing controversies exist about the reality and properties of
the respective planets.  Some of those systems require further studies
on their composition and the properties of the various components.
In general, the identified systems
may host additional planets, noting that for some of those systems, as, e.g.,
the Alpha Centauri system, evidence of still unconfirmed planets already
surfaced \citep[e.g.,][]{wagn21}.  In case of 2M~J0441~+~2301,
it is still unclear if one of the system components is a low-mass brown dwarf
or a giant planet \citep{todo10,bowl15}.  Other objects of interest include
HD~188753 and Gliese~667.  Regarding the systems Fomalhaut and HD~131399,
solid evidence has been provided that the previously proposed planets do not
exist.  Therefore, in the context of this study, these systems are considered
as retracted and are thus not listed as controversial.

Following \cite{kona05}, the system of HD~188753 contains a solar-type star
(with a temperature and mass akin to the Sun) that is host to a hot Jupiter
as well as a close binary; hence, the classification of a planet-hosting triple
stellar system.  However, follow-up observations by \cite{egge07} based
on Doppler measurements did not confirm the planet's existence;
therefore, we listed this system as controversial.
Gliese~667 (or GJ~667) is another case of a long-standing controversy.
GJ~667 is a planet-hosting triple star system noting that GJ~667C, the
least massive of the three components, is homestead to two confirmed
super-Earth planets \citep{fero14,robe14}.  Previously, it was argued that
GJ~667 would host several additional planets, including three planets
in its HZ \citep{angl12,angl13,delf13}.  Subsequent estimates based on
the Bayesian analysis of radial velocity data \citep{fero14} reduced that
number to two (or, less likely, three).  A key aspect in the determination
of the correct number of planets hosted by M-dwarfs such as GJ~667C is
the adequate analysis of stellar activity, which is particularly challenging
\citep[e.g.,][and references therein]{tuom19,lafa21}.

Other cases of ongoing discussions and controversies include the systems
of HD~4113, HW~Virginis, KIC~7177553, and 40~Eridani.
According to \cite{chee18}, HD~4113 is a complex
dynamical system consisting of a giant planet, stellar host and a known
M-dwarf companion; see also \cite{mugr14} and \cite{mugr19} for additional
results.  The system also contains an ultracool substellar companion
of late-T spectral type; its estimated mass given as about 66 $M_{\rm J}$
is evidently beyond the Jovian planet limit.  Evidence
of the planet Ab has been reported by \cite{tamu08}.  However, open questions
about the system's composition remain as the memberships of some
components are not fully established.
In fact, one of the system components is situated at an angular separation
of 43$^{\prime\prime}$, corresponding to a projected separation of
2000~au\footnote{Noting that HD~4113 may be a planet-hosting
binary, HD~4113b, albeit undisputed, is placed in the unconfirmed
category of Table~7 and subsequent representations.}.

HW~Virginis is a system
that is considered potentially unstable; it appears to consist of two stars,
a Jupiter-type planet, and a brown dwarf; thus, a triple star system.
Previous work on the stability of that system was pursued by \cite{beue12}
and \cite{horn12}.  KIC~7177553 is a young system observed by the
{\it Kepler} satellite \citep{lehm16}.  It consists of a pair of binary stars;
hence, a quadruple system.  The system also contains a super-massive
Jupiter-type planet (but below the brown dwarf mass limit).  However,
considering the system's complexity and dynamics, there is a need for
additional transit data to ultimately confirm the planet's existence.

For 40~Eridani, \cite{ma18} reported a super-Earth in a 42.4~d orbit,
using data sets taken by multiple spectrographs, including HIRES.
Considering the close proximity of this system to Earth, given as 5.04~pc
(see Table~1), this would have made that planet-hosting triple star system
the second closest to Earth known to date, only topped by Proxima Centauri.
However, in 2021 \citeauthor{ros21} determine a significant periodicity at
42 days in the HIRES $S$-value measurements, thus concluding that 42 days
is the likely rotation period of that star.  Moreover, \cite{ros21}
identified evidence of a long-period magnetic activity cycle with the
nearly same period.  Hence, the previously acclaimed planet is
most likely a false positive.


\subsection{Refutations and Retractions}

\subsubsection{Fomalhaut~b}

Following \cite{kala08}, and references therein, Fomalhaut ($\alpha$~PsA)
is a triple stellar system, with its main component, an A-type star,
shrouded by an extended disk structure, indicating ongoing planet formation.
According to this study, optical observations point to an exoplanet candidate,
Fomalhaut~b.  Moreover, optical dynamical models of the interaction
between the putative planet and the gas and dust belt indicate that
Fomalhaut~b's mass is at most three times that of Jupiter.  Open questions
about observational features at different wavelengths (mm-regime),
including variabilities, however remained --- albeit work by \cite{curr12}
provided support for the interpretation of \citeauthor{kala08}

However, \cite{lawl15} in a follow-up study pointed out that although the
planet candidate Fomalhaut~b is bright in optical light but undetected in
longer wavelengths, requiring a large, reflective dust cloud. According
to the authors, the new observations point to an extremely eccentric
orbit ($e \simeq 0.8$), indicating that the existence of Fomalhaut~b
is not consistent with the system's eccentric debris ring.  Hence, an
irregular satellite swarm around a super-Earth is proposed instead.
\cite{lawl15} also predicted that the feature mimicking the existence of
Fomalhaut~b is expected to expand until it either dissolves or becomes
too faint to be seen.

This view is in alignment with a recent study by \cite{gasp20}, based on
HST observations, which arrived at a different conclusion than previously
conveyed by \cite{kala08}.  \cite{gasp20} revisited published data in
together with more recent HST data, concluding that the source is likely
on a radial trajectory, evidently related to a collision between two large
planetesimals, and has started to fade away.  Apparently, we are witnessing
the effects of gravitational stirring due to the orbital evolution of
hypothetical planet(s), which may still exist in that system.  However,
the exoplanet Fomalhaut~b does not exist.

\subsubsection{HD~131399~Ab}

HD~131399, located in the constellation of Centaurus, was previously classified
as a planet-hosting multiple stellar system.  However, this assessment has changed.
As a consequence, HD~131399 is not any further considered in this study.  Earlier,
\cite{wagn16} reported the discovery of a Jovian exoplanet, named HD~131399~Ab,
in that system based on direct imaging.  They classified HD~131399~Ab as a Jovian
planet of $4 \pm 1$~$M_{\rm J}$ with a relative wide orbit situated in a hierarchic
triple star system.  \cite{wagn16} already pointed out that the planetary orbit is
potentially unstable.  Additional details about the system were given by
\cite{lagr17}, who also noted that the planet's existence challenges
conventional planet formation theories.  The object was originally recorded
using the SPHERE imager at ESA's Very Large Telescope.

However, a follow-up study by \cite{niel17} indicated that the candidate planet
is, in actuality, a background star instead.  The authors conveyed results
from JHK1L' photometry and spectroscopy obtained with the Gemini Planet Imager,
VLT/SPHERE, and Keck/NIRC2, as well as a reanalysis of the previous analysis data.
This finding is also consistent with the object's proper motion.  Furthermore, as
discussed by \cite{niel17}, if HD~131399~Ab were a physically associated object,
its projected velocity would exceed the escape velocity given the mass and distance
to HD~131399~A.  Hence, the putative exoplanet is most likely a background K or M
dwarf instead.  \citeauthor{wagn16} now concur with most if not all of these
findings, and on April 14, 2022 they retracted their article.

\section{Statistical Analysis and Interpretation}

\subsection{Triple Systems}

\subsubsection{Stellar Components}



Table 8 provides information on the planet host stars of confirmed triple stellar systems;
see also Fig.~8 for additional information.  There is a notable range
regarding spectral type, with the upper end given by Kepler-13A, an A-type star
\citep{shpo14}.  Based on their report, the main stellar parameters are given as
$T_{\rm eff} = 7650 \pm 250$~K, $M_\star = 1.72 \pm 0.10$~$M_\odot$, and 
$R_\star = 1.71 \pm 0.40$~$R_\odot$.  In fact, the majority of planet host stars
are G dwarf noting that 16~Cygni~B most closely resembles the Sun
\citep[e.g.,][]{metc15}.  The least massive star is Proxima Centauri with a mass
of 0.12~$M_\odot$ (see Sect.~4 for details).  In Table 8, for each host star
information is given about the respective stellar effective temperature, mass,
radius, and  luminosity as obtained from the literature.

Various kinds of techniques have been employed for deriving the acquired information,
which in regard to the mass also considered stellar evolution scenarios.
Stellar surface temperatures have typically been deduced spectroscopically.
Since the stellar effective temperature, radius, and the stellar luminosity are
not independent of one another, extra opportunities exist for deriving the missing
parameter.  In case of the systems HAT-P-8, K2-290, Kepler-444, and KOI-5, the
star's luminosity value has been deduced from the stellar effective temperature
and radius.  For some main-sequence stars, the mass--luminosity relation proved
useful as well; see, e.g., \cite{cunt18} for information on the $M-L$ relationship
for K dwarfs.

The masses of the planet host stars of confirmed triple stellar systems range
from 0.12~$M_\odot$ (Proxima Centauri) to 1.72~$M_\odot$ (Kepler-13A).
Regarding the Kepler-13AB system, valuable information has been obtained by
\cite{howe19} based on the high-resolution imaging instrument, {\'A}lopeke,
at the Gemini-N telescope indicating that Kepler-13b is a highly irradiated
gas giant with a bloated atmosphere.  The denoted mass range
can be compared to previous work by, e.g., \cite{mugr19} who studied
the properties of planet host stars, which are typically not members of multiple
stellar systems, while also using information provided by {\it Gaia} astrometry.
They found that those stars exhibit masses in the range between about
0.078 $M_\odot$ and 1.4 $M_\odot$ with a peak in their mass distribution between
0.15 $M_\odot$ and 0.3 $M_\odot$.  However, the planet host stars in triple
systems are identified to be more massive; see Sect.~5 for further comments.

Important sources for accurate information about stellar diameters and temperatures
of the various stellar components
have been given by \cite{boya12,boya13} and \cite{whit13}.  They conveyed interferometric
angular diameter measurements made with the CHARA Array for a notable range of
main-sequence stars.  These data were used in combination with the {\it Hipparcos}
parallaxes and new measurements of the stellar bolometric fluxes to compute
absolute luminosities, linear radii, and stellar effective temperatures.
Previously, \cite{bell09} determined linear radii and effective temperatures for
numerous exoplanet host stars based, in part, on qualified estimates of
bolometric fluxes and reddenings through spectral-energy distribution fits.
Improved spectroscopic parameters for various transiting planet host stars
have been given by \cite{torr12}.  Relevant information for most objects
has also been obtained through the {\it Gaia} data release \citep{brow18}.

In case of late-K and M dwarfs, detailed studies
by \cite{mann15} led to improved empirical measurements of the stellar
effective temperatures, bolometric luminosities, masses, and radii.
Another important contribution has been made by \cite{bain18} who derived
fundamental properties of various planet host stars based on angular diameter
measurements from the NPOI.
Another source of updated stellar parameters is the work of \cite{rain20} who obtained accurate
angular diameters for many southern stars based on the PIONIER beam combiner
located at the VLTI.  Significant credit should also be extended to the {\it Kepler}, K2, and
TESS missions for providing valuable stellar and planetary information.
In many other cases, valuable stellar parameter determinations were
provided in conjunction with the respective planetary discoveries (see Table~1) or in subsequent
work with the effort of constraining the relevant parameters of the star-planet systems.


\subsubsection{Planetary Components}






As conveyed in Table 1, 3, and 5, a total of 27 confirmed planet-hosting triple star
systems has been identified, and the total number of confirmed planets is 38.
Various detection methods have been utilized, while noting the most successful
method has been the Radial Velocity method, yielding 21 planets, followed by the
Transit method, utilized by {\it Kepler}.  Minor contributions were made through
direct imaging and by astrometry and eclipse timing.  Both gas giants (the dominant type)
and terrestrial planets (including super-Earths) have been identified.  Thanks to the
Radial Velocity method, 24 Jupiter-type planets have been discovered.

In addition, a small number of Earth-mass and sub-Earth-mass planets have been found as well,
located in the systems of Alpha Centauri, Kepler-444, and LTT 1445; see \cite{fari22},
\cite{camp15}, and \cite{wint19}, respectively; see Table 9 and 10 for details and further
information.  The exoplanet of lowest mass, which is 0.034 $M_\oplus$, has been found
within the multiple planetary system hosted by Kepler-444 \citep{camp15}.
Kepler-444 is identified as a metal-poor Sun-like star from the old population of
the Galactic thick disk.

Despite the fact that the total number of planets hosted by triple star systems is
relatively small, an attempt can be made to identify some general trends.  In Fig.~9, we
display of the planetary eccentricity $e_{\rm p}$ as a function of the semimajor axis $a_{\rm p}$;
here we also distinguish between planets of single and multi-planetary systems.
Furthermore, in Fig.~10, we report the planetary masses $M_{\rm p}$ again as a
function of the semimajor axis $a_{\rm p}$.  Note that an arrow was used when only
the planet's minimum or maximum mass is known.  (The former case typically occurs when
a planet has been identified through the Radial Velocity method.)

Both Fig.~9 and 10 are, in essence, scatter plots with no discernible trend information.
Most planets have low or moderately-high eccentricities, and there are two planets,
which are 16~Cygni~Bb and HD~41004Ab, with eccentricities exceeding 0.65.  Very small
eccentricities (i.e., close to or indistinguishable from zero) can be found
for close-in planets, a result also known from previous studies.  Examples of 
early work include contributions by \cite{marc04} and \cite{butl06}, with the former
being a Doppler planet survey of 1330 FGKM stars.  By making use of the Lick, Keck,
and ATT, 75 planets have been found which broadly speaking exhibit a very similar
$e_{\rm p} - a_{\rm p}$ distribution as depicted in Fig.~9.

Similar results have been found by \cite{butl06} who presented a catalog of nearby
exoplanets.  They examined 168 planets regarding their $e_{\rm p}$ and $a_{\rm p}$
properties; in this case, 11 planets have been identified with eccentricities in
excess of 0.65.  However, no viable conclusion about the abundance of high-eccentricity
planets in triple stellar systems can be conveyed considering the inherent limitations
of small number statistics.  Our results are also consistent with work by \cite{kane12}
who studied the exoplanet eccentricity distribution from {\it Kepler} planet candidates.
Among other results, they found that the mean eccentricity of the {\it Kepler} candidates
decreases with decreasing planet size indicating that smaller planets are preferentially
found in low-eccentricity orbits.  In our case, those planets happen to be close-in
planets.

In Figure 11, we relate the planetary mass to the mass of the planet host star.
Here it is found that planets of small masses, i.e., Earth-mass and sub-Earth-mass planets
are preferably found in the environments of low-mass stars, i.e., K and M dwarfs.
Those planets also occur within multi-planetary systems, notably the systems hosted
by Proxima Centauri and Kepler-444A; however, additional observations of
planets in tertiary stellar systems would be required to allow conclusions
of statistical significance.  Previous results on the occurrence and mass
distributions of close-in Super-Earths, Neptunes, and Jupiters have been given by
\cite{howa10} and others, which should be considered as part of future analyses.


\subsection{Quadruple Systems}

So far, three systems with (at least) four stellar components have been identified
that are hosts to confirmed or suspected exoplanets.  They are: 30~Arietis,
Kepler-64, and KIC~7177553.  The most convincing case is Kepler-64, whereas
for the systems of 30~Arieti and KIC~7177553 notable uncertainties exist.

Regarding Kepler-64, a planet has been found by \cite{schw13} constituting a
transiting circumbinary planet around an eclipsing binary in the {\it Kepler} field.
According to that report, the planet was discovered by volunteers as part of the
Planet Hunters citizen science project.  The physical and orbital parameters
of both the host stars and planet were obtained via a photometric-dynamical model,
simultaneously fitting both the measured radial velocities and the {\it Kepler} light curve
of the host objects.  According to \cite{schw13}, the $6.18 \pm 0.17$~$R_\oplus$
planet orbits outside the 20 day orbit of an eclipsing binary consisting of an
F dwarf ($1.734 \pm 0.044$~$R_\odot$, $1.528 \pm 0.087$~$M_\odot$) and M dwarf
($0.378 \pm 0.023$~$R_\odot$, $0.408 \pm 0.024$~$M_\odot$).  The authors report
an upper planetary mass limit of 0.531~$M_{\rm J}$.  Outside the planet's orbit,
at $\sim$1000~au, a previously unknown visual binary has been identified \citep{schw13}
that is likely bound to the planetary system, making this the first known case
of a planet-hosting quadruple star system, a so-called ``double-double".

30~Arietis and KIC~7177553 are two uncertain cases of planet-hosting quadruple
systems; the putative planets have been identified by \cite{guen09} and \cite{lehm16},
respectively.  30~Arietis constitutes either a planet-hosting stellar quadruple
system or a quintuple stellar system without a confirmed planet.  According to
\cite{robe15}, 30 Arietis A and B are separated by 1,500~au.  The two stellar
components are at almost the same distance and have very similar proper motions;
thus, they are almost certain gravitationally bound.  Furthermore, the main
components of both systems are both binaries with a composite spectra indicative
of F-type stars.  Following \cite{morb74}, 30 Arietis A is a spectroscopic binary
with an orbital period of 1.1 days.  Based on work by \cite{robe15} and \cite{guen09},
30 Arietis B is categorized as a red dwarf companion at a distance of 22~au;
moreover, there is another object, denoted as Bb, at about 1~au.  \cite{guen09}
previously identified 30~Arietis~Bb as a planet.  However, recently measurements
of the planetary orbit by \cite{kief21} led to evidence that the object might fall
in the mass range of a brown dwarf or red dwarf instead.

According to \cite{lehm16}, KIC~7177553 is a system previously observed
by the {\it Kepler} satellite.  It consists of a pair of binary stars;
hence, a quadruple system.  The system also contains a super-massive
Jupiter-type planet, but apparently below the brown dwarf mass limit. 
That latter object was revealed through eclipse timing variations; it has
a period of approximately 529 days.  Based on the work by \cite{lehm16}
employing RV measurements, it became obvious that the same {\it Kepler}
target contains another eccentric binary, which is on a 16.5-day orbital
period.  \cite{lehm16} also pointed out that the separation distance of the
two binaries is about 167~au, and that they have nearly the same magnitude
(to within 2\%).  In addition, the close angular proximity of the two binaries
and very similar $\gamma$-velocities strongly suggest that KIC~7177553 is one
of the rare planet-hosting double-double systems where at least one system
is eclipsing.  Moreover, both systems consist of slowly rotating, nonevolved,
solar-like stars of comparable masses.  However, considering the system's
complexity and dynamics, there is a need for additional transit data to
ultimately confirm the planet's existence.


\section{Case Study: The Alpha and Proxima Centauri System}

\subsection{Basic Properties}

Alpha Centauri is the closest star (if reduced to one component) and the closest stellar system
to Sun-Earth.  Hence, any planet hosted by that system constitutes the closest exoplanet to Earth,
regardless of any future planet detections.  Alpha Centauri is a hierarchic triple stellar system,
consisting of a quasi-binary, i.e., star A and B, of spectral type G2~V and K1~V, respectively
\citep{torr06}, and a distant component, which is Proxima Centauri, an M dwarf.
Alpha Centauri A and B have effective temperatures of 5790~K and 5260~K and masses
of 1.13 $M_\odot$ and 0.97 $M_\odot$, respectively; see \cite{thev02} and \cite{pour16}.
For additional information on stellar parameters see, e.g., \cite{boya13}, \cite{kerv17a},
and \cite{akes21}.

Alpha Cen~A and B are at a distance of 1.339~pc from the Sun, whereas Proxima Cen
is at a distance of 1.302~pc \citep{brow18}; these values have been obtained by {\it Gaia}.
The pair Alpha Cen~A and B is in a highly eccentric orbit with their relative distance changing
between 11.2~au and 35.6~au; the orbital period is given as 79.91 years \citep{hart08}.
The system's age has been determined as $5.3 \pm 0.3$~Gyr \citep{joyc18} based
on classically and asteroseismologically constraint stellar evolution models.

Proxima Centauri, also known as Alpha Centauri~C or GJ~551, is $8200_{-300}^{+400}$~au
apart (semimajor axis) from Alpha Cen~A and B.  It has an orbital period of
$511,000_{-30,000}^{+41,000}$ years \citep{akes21}; see also \cite{kerv17b}.
Furthermore, Proxima Cen is on a highly eccentric orbit; its current distance from Alpha Cen~AB is
12,950~au --- a number that is poised to change as time progresses.  Proxima Cen is
a red dwarf of spectral type M5.5~Ve \citep{bess91}.  Its stellar effective temperature is about
3000~K; moreover, its stellar mass and radius are given as about 0.12~$M_\odot$
and 0.15~$R_\odot$ \citep{delf00,segr03,boya12,mann15,riba17}, respectively.  Proxima Cen
also exhibits significant amounts of stellar activity --- comprised of emission, flares and
outflows detectable in different wavelength regimes \citep[e.g.,][]{fuhr11,ayre14,robe16,pavl17,howa18}.
Note that these features are highly relevant for its prospects of circumstellar habitability.


\subsection{Planet Detections}

For Alpha Centauri A and B, there are no confirmed planet detections --- yet, regarding Alpha Cen~A,
there is tentative evidence in favor of a possible planet or exozodiacal disk as discussed by
\cite{wagn21}; see, e.g., \cite{wang22} for additional analyses.  A previous report about an Earth-mass
planet hosted by Alpha Cen~B by \cite{dumu12} has been superseded by \cite{rajp16} who demonstrated
that this earlier detection was a false positive.  On the other hand, Proxima Centauri is known
to harbor three planets, discovered between 2016 and 2022 by \cite{angl16}, \cite{dama20}, and
\cite{fari22}, respectively.  Facilities that made these observations possible include La Silla,
VLT, and HST.  Theoretical results indicating the possible existence of planets in the Alpha Cen
system have been obtained by \cite{quin02} who examined terrestrial planet formation for
different dynamic scenarios.  General studies on planet detectability in the Alpha Cen system have
been published by \cite{zhao18}.

\cite{angl16} identified an Earth-mass planet (minimum mass 1.3~$M_\oplus$) in orbit about Proxima Cen
with a period of about 11.2~d at a semimajor-axis distance of close to 0.05~au.  Their analysis
indicated that this object is situated in the stellar habitable zone, presumably permitting liquid water on the
planetary surface.  The measurement allowing for the detection of the planets were performed using two spectrographs,
which are the HARPS on ESO's 3.6 m Telescope at La Silla Observatory and UVES on the 8 m Very Large Telescope
at Paranal Observatory.  In 2020, \citeauthor{dama20} identified based on the Radial Velocity method
a low-mass planet hosted by Proxima Cen at a distance of 1.5~au.  According to its minimum mass of 5.8~$M_\oplus$,
this planet is categorized as either a large super-Earth or a mini-Neptune.  However, it is situated well beyond
the outer limit of the stellar habitable zone.  The planet's detection is partially based on ESO's HARPS
instrument, and in 2020, the planet's existence was ultimately confirmed by Hubble astrometry data from 1995;
see \cite{bene99} and references therein.

A third planet hosted by Proxima Cen was discovered by \cite{fari22}.  This work is based on a re-analysis of
observations taken with the ESPRESSO spectrograph at the VLT aimed at evaluating the presence of a third
low-mass planetary companion, which according to the authors started emerging during a previous campaign. 
\cite{fari22} found a signal at $5.12 \pm 0.04$~d with a low semi-amplitude.  According to
the authors, the analysis of subsets of the ESPRESSO data, the activity indicators, and chromatic RVs suggest
that this signal is not caused by stellar variability but instead by a planetary companion with a minimum mass
of $0.26 \pm 0.05$~$M_\oplus$ (about twice the mass of Mars) orbiting at 0.029~au from its host.
Following \cite{fari22}, the planet's orbital eccentricity is compatible with a circular orbit. 
This observation is of pivotal importance as it indicates a planet with a sub-Earth mass is the immediate
Sun-Earth neighborhood.


\subsection{Orbital Stability and Habitability}

There is little doubt about the orbital stability of the stellar components and the Alpha Centauri
system planets.  This view has been confirmed by \cite{boyl21} who investigated the orbital stability\footnote{
Due to the highly hierarchic nature of the triple star systems studied here, including Alpha Centauri, the
binary approximation in orbital stability simulations is often appropriate; see work by, e.g.,
\cite{holm99}, \cite{fatu06}, and \cite{cunt07}, among other contributions.} of the outlying stellar component
and the two heretofore identified exoplanets based on the Hill stability criterion.  Examples of previous
work on planetary habitability for the Alpha Centauri system, such as the circumstellar environment of
Proxima Centauri, include studies by \cite{barn16} and \cite{mead18}.

\cite{mead18} used 1-D coupled climate-photochemical models to generate self-consistent atmospheres for
several evolutionary scenarios, including Earth-like atmospheres, with both oxic and anoxic compositions.
They showed that these modeled environments can be habitable or uninhabitable at Proxima Cen b's position
in the habitable zone.  These results (according to the authors) are applicable not only
to Proxima Cen b, but to other terrestrial planets orbiting M dwarfs as well --- especially in consideration
of that M dwarfs are disadvantageous for hosting planets with life compared to, e.g., K dwarfs \citep[e.g.,][]{cunt16}.
On a perhaps less serious note, \cite{mari18} calculated, through employing multi-parameter Monte Carlo simulations,
the minimal crew for a multi-generational space journey toward Proxima b, estimated to last about 6300~years (their
primary model).  This work re-emphasizes the impossibility of interstellar travel under realistic assumptions based
on current technology.



\section{Additional Analyses}



Another item of interest is the statistical analysis of the planet host stars in triple stellar systems.
Thus, we focus on the 27 confirmed systems, defined as that all three stellar system components and
at least one planet have been verified (see Table 5 and 6).  Table 8 shows that those stars encompass
a large range of masses, including M dwarfs such as Proxima Centauri, Gliese 667C, and LTT~1445A with
masses as low as 0.12~$M_\odot$; see \cite{riba17}, \cite{toko08}, and \cite{wint22}, respectively.
On the other hand, planet-hosting stars of higher mass have been identified as well with 51~Eri~A,
an F-type star of mass $1.75 \pm 0.05~M_\odot$ \citep{simo11} as the current number one.

Next we focus on the distribution of stellar spectral types (see Fig.~8, and Table 6 and 11).  It is found that
the system's main components are relatively more massive compared to main-sequence stars on average.
Specifically, stellar primaries tend to be more massive than typical main-sequence single stars,
a finding also valid for stellar secondaries though less pronounced.  Tertiary stellar components
are almost exclusively low-mass stars of spectral type M. Hence, in regard to planetary host stars,
there is an increased abundance of A, F, and G-type stars.

In Table~12, we offer a more detailed statistics for stellar spectral types using a number scheme
such as that A0, F0, G0, K0 and M0 stars are denoted as 1.0, 2.0, 3.0, 4.0, and 5.0, respectively.
Stars of intermediate spectral types are denoted accordingly.  Furthermore, we refrain from considering
non-main-sequence stars, which led to the dismissal of 91~Aqr~A, HD~185269~A, and Psi$^1$~Dra~A as
those stars are further evolved (with Psi$^1$~Dra~A borderline); see \cite{bain18}, \cite{mout06},
and \cite{endl16}, respectively.  In most other cases, the spectral types and subtypes are well-known.
If missing, this information has been deduced based on the stellar effective temperature using the
scale of \cite{gray05}.

Our analysis shows that at present the average spectral type of planet host stars in triple stellar systems
is given as G3~V.  Furthermore, there is no statistically significant difference whether or not
the system harbors close-in planets or the planets are situated further out; here a divider at 0.1~au
(semimajor axis) is used.  We also compared this result to planet-hosting binaries, applied to S-type
systems, where the planet is in orbit about one stellar component --- instead of both, with that latter
case referred to as P-type system; see, e.g., \cite{dvor82}.  The data for the binary systems as considered
are taken from \cite{pila19}.  In this case, the average spectral type of planet host stars is given as G8~V.
Evidently, those stars are of later type; however, the difference to planet-hosting stars situated in
triple star systems is about one standard deviation.  Hence, a larger data base, particularly concerning
triple star systems would be advantageous to further explore that statistical trend.

This statistical outcome can also be compared with the general spectral distribution for main-sequence
stars; the latter is in part informed by \cite{krou01,krou02} and \cite{chab03} who pursued
detailed analyses of the universal stellar initial mass function for the various components of the
Milky Way.  These studies reconfirmed that the mass function is heavily skewed toward the low end
of the main-sequence.  Previous work by \cite{ledr01} for the spectral types of main-sequence stars
located in the solar neighborhood identified the following distribution:
A: 0.6\%, F: 3.0\%, G: 7.6\%, K: 12.1\%, and M: 76.5\%.
(This also means that the Sun constitutes a high-mass star, although this attribution is barely made.)

These percentages are strikingly similar to the result obtained by
E. Mamajek in 2016, unpublished (used with permission)\footnote{
https://figshare.com/articles/figure/Fraction{\_}of{\_}Stars{\_}by{\_}Spectral{\_}Type{\_}in{\_}the{\_}Solar{\_}Vicinity/3206527},
which for the various spectral types reads: A: 0.6\%, F: 3\%, G: 6\%, K: 13\%, and M: 72\%
(while ignoring a normalization factor of 1.057).  Hence, the typical spectral type of
general main-sequence stars is about M3 (the median), which is much closer to the lower end of the main-sequence
than found for planet host stars in triple stellar systems.  Consequently, from a statistical point of view, the
latter types of stars are more massive, concurrent with a notably higher surface temperature and luminosity,
than main-sequence stars in the solar neighborhood.


\section{Summary and Conclusions}

The focus of this study is to present a catalog, including background
information and associated interpretation, of planet-hosting triple
and quadruple star systems by making use of observations and data
available in the literature.  We explored statistical properties
of those kinds of systems with a focus on both the stars and the
planets.  So far, about 30 triple systems and one confirmed
quadruple system have been identified.  Regarding both the
triple and quadruple systems, there is also a small number of
controversial cases, which have been discussed as well.

The total number of planets identified in planet-hosting triple
and quadruple stellar systems is close to 40.  All triple star systems
are highly hierarchic, consisting of a quasi-binary complemented by
a distant stellar component, which is in orbit about the common center
of mass.  Furthermore, the confirmed and unconfirmed planet-hosting quadruple systems
are, in fact, pairs of close binaries (``double-doubles"), with one of
the binaries harboring an exoplanet.

We acknowledge that planet-hosting systems of higher multiplicity are
expected as well.  Actually, there may be a smooth transition between
high-order multiple stellar systems and gravitationally bound open clusters
containing a high number of stellar components.  Owing to the complex patterns
of gravitation interaction in those systems, possible exoplanets hosted in
high-order systems may however be restricted to relatively tight S-type
configurations and to P-type configuration pertaining to quasi-binaries.
But so far no detections for systems of five or more stars have been
obtained.

We identified the following properties regarding planet-hosting
triple star systems:

\medskip\noindent
(1) The closest planet-hosting multiple star system is Alpha \& Proxima Centauri
at a distance of 1.302~pc.  Previously, this system has received heightened attention
regarding both orbital stability and habitability; some of these aspects are discussed
in this study.

\medskip\noindent
(2) Planet-hosting multiple star systems, including a small
number of unconfirmed cases, are at present found in 22 of the 88 constellations.
As expected, a high number of systems exists at or near the Summer Triangle, 
largely owing to the contributions from the {\it Kepler} mission.

\medskip\noindent
(3) All or almost all planets are found to be in orbit about single stars rather than
pairs of stars; these settings are readily referred to as S-type orbits.  So far,
no planets have been identified that are in orbit about three stars at once
(``circumtrinary" planets).  The presence of those planets is incompatible with
highly hierarchical triple star configurations.

\medskip\noindent
(4) Besides the various systems with one identified planet, there are at present
seven multi-planetary systems hosted by triple star systems; the total number of planets
in those systems is 18.  The highest number of planets have been found in the system
of Kepler-444, which are all identified as sub-Earth--type planets.

\newpage

\noindent
(5) The dominant method of planet detections has been the Radial Velocity method,
yielding 21 confirmed planets in triple star systems. The second most important
method has been the Transit method, utilized by {\it Kepler}.  Minor contributions
were made through direct imaging and by astrometry and eclipse timing.

\medskip\noindent
(6) Both gas giants (the dominant type) and terrestrial planets (including super-Earths)
have been identified.  Thanks to the Radial Velocity method, 24 Jupiter-type planets were discovered
in triple stellar systems.  Moreover, a small number of Earth-mass and sub-Earth-mass planets
have been detected as well --- including planets hosted by Proxima Centauri prompting a
considerable array of studies about planetary formation, orbital stability, and habitability.

\medskip\noindent
(7) The data show that almost all stars are main-sequence stars, as expected.  However,
the stellar primaries tend to be more massive (i.e., corresponding to spectral types
A, F, and G) than expected from single star statistics, a finding also valid for
stellar secondaries but less pronounced.  Furthermore, tertiary stellar components
are almost exclusively M dwarfs --- a finding also mirrored by the statistical distribution
of stellar spectral types in the solar neighborhood \citep[e.g.,][]{krou01,krou02,chab03}.

\medskip\noindent
(8)  Our analysis also shows that for the average spectral type of planet host stars
in triple stellar systems is given as G3~V and, furthermore, there is no identifiable difference
whether or not the system harbors close-in or far-away planets.  Thus,
planet host stars tend to be more massive and, consequently, have a notably higher
surface temperature and luminosity, than typical main-sequence stars (as defined
by the median of the stellar mass sequence).

\bigskip

Topics of future studies pertaining to planets in higher order systems are expected to
include studies of orbital stability, history of formation, and habitability ---
particularly regarding Earth-type planets and Super-Earths.
Previous studies about the influence of stellar multiplicity on planet formation
have been given by, e.g., \cite{wang14}, \cite{dutr14}, and subsequent work, indicating that
planet formation due to stellar companions is noticeably reduced but nonetheless
substantial.  It would be of great interest to expand these investigations while
taking into account expanded parameter domains including effects associated with
various types of outside forcings.

Previous work about the stability of planets in triple and higher order stellar systems have
been given by, e.g., \cite{verr07}, \cite{hame15}, \cite{corr16}, \cite{buse18}, and \cite{myll18}.
However, these studies are typically limited to triple stellar systems, especially to those of
highly hierarchic nature --- an approach well motivated by the current observations.
On the other hand, in consideration of ongoing and future observational campaigns, more
intricate compositional and dynamical star-planet structures are expected to be found.  Those
findings are expected to inform future studies of orbital stability and habitability.  The
latter will need to take into account both gravitational and radiative boundary conditions
imposed by the various system components, which in turn will require additional observational
and theoretical studies.  Other contributions are expected to arise from the development
of future space mission concepts as, e.g., LUVOIR \citep{luvo19}, HabEx \citep{gaud20},
and LIFE \citep{quan21}, which are poised to provide additional crucial insights into the physics of
multiple stellar and multiple planetary systems.



\begin{acknowledgments}
This work has been supported by the Department of Physics, University
of Texas at Arlington.  The authors acknowledge comments by an anonymous
referee.  Note that data from a large variety of sources available in the
literature have been considered in support of this study as discussed in the
text.  Early results associated with this study were previously presented by G. E. L.
as part of a Honors Thesis submitted to UTA's Honors College.  Moreover,
the authors appreciate comments by Z. E. Musielak on an early version
of the manuscript.
\end{acknowledgments}

\newpage

\appendix

\section{SUMMARY OF ACRONYMS}

\begin{tabular}{ll}
\noalign{\smallskip}
\hline
\noalign{\smallskip}
Acronym       &  Meaning \\
\noalign{\smallskip}
\hline
\noalign{\smallskip}
   ALMA       & Atacama Large Millimeter/submillimeter Array \\
   ATT        & Anglo-Australian Telescope \\
   CHARA      & Center for High Angular Resolution Astronomy \\
   CNUO       & Chungbuk National University Observatory \\
   ESA        & European Space Agency \\
   ESO        & European Southern Observatory \\
   ESPRESSO   & Echelle Spectrograph for Rocky Exoplanet- and Stable Spectroscopic Observations \\
   GJ         & Gliese-Jahrei{\ss} catalogue \\
   HabEX      & Habitable Exoplanet Observatory \\
   HARPS      & High Accuracy Radial velocity Planet Searcher \\
   HAT        & Hungarian-made Automated Telescope \\
   HD         & Henry Draper catalogue \\
   HIRES      & High Resolution Echelle Spectrometer \\
   HST        & Hubble Space Telescope \\
   HZ         & Habitable Zone \\
   IAU        & International Astronomical Union \\
   KIC        & Kepler Input Catalog \\
   KOI        & Kepler Object of Interest \\
   K2         & Kepler-2 \\
   LIFE       & Large Interferometer for Exoplanets \\
   LTT        & Luyten Two-Tenths catalog \\
   LUVOIR     & Large UV/Optical/IR Surveyor \\
   NASA       & National Aeronautics and Space Administration \\
   NIRC2      & Near Infrared Camera 2 \\
   NPOI       & Navy Precision Optical Interferometer \\
   PIONIER    & Precision Integrated-Optics Near-infrared Imaging ExpeRiment \\
   RA         & Right Ascension \\
   RV         & Radial Velocity \\
   SIMBAD     & Set of Identifications, Measurements and Bibliography for Astronomical Data \\
   SOAO       & Sobaeksan Optical Astronomical Observatory \\
   SPHERE     & Spectro-Polarimetric High-contrast Exoplanet Research \\
   TESS       & Transiting Exoplanet Survey Satellite \\
   TIC        & TESS Input Catalog \\
   UTA        & University of Texas at Arlington \\
   UVES       & UV-Visual Echelle Spectrograph \\
   VLT        & Very Large Telescope \\
   VLTI       & Very Large Telescope Interferometer \\
   WASP       & Wide Angle Search for Planets \\
\noalign{\smallskip}
\hline
\noalign{\smallskip}
\end{tabular}


\vfill
\eject




\begin{figure}[htb]
\begin{center}
\includegraphics[scale=0.75]{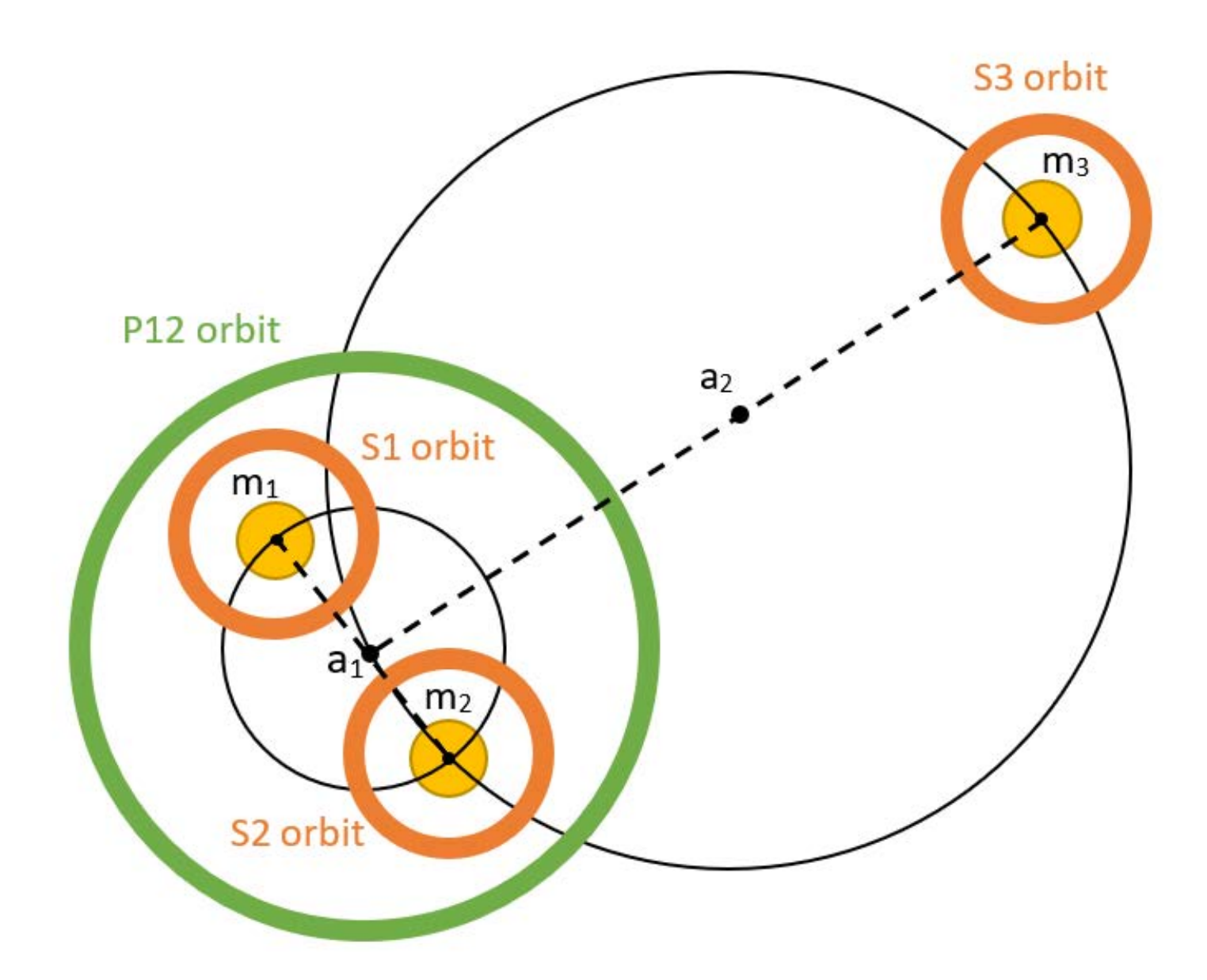}
\caption{
Possible orbits of a planet in an hierarchic triple star system.
Note that $m_1$, $m_2$, and $m_3$ do not necessarily agree.
\label{fig:1}}
\end{center}
\end{figure}

\newpage

\begin{figure}[htb]
\begin{center}
\includegraphics[scale=0.35]{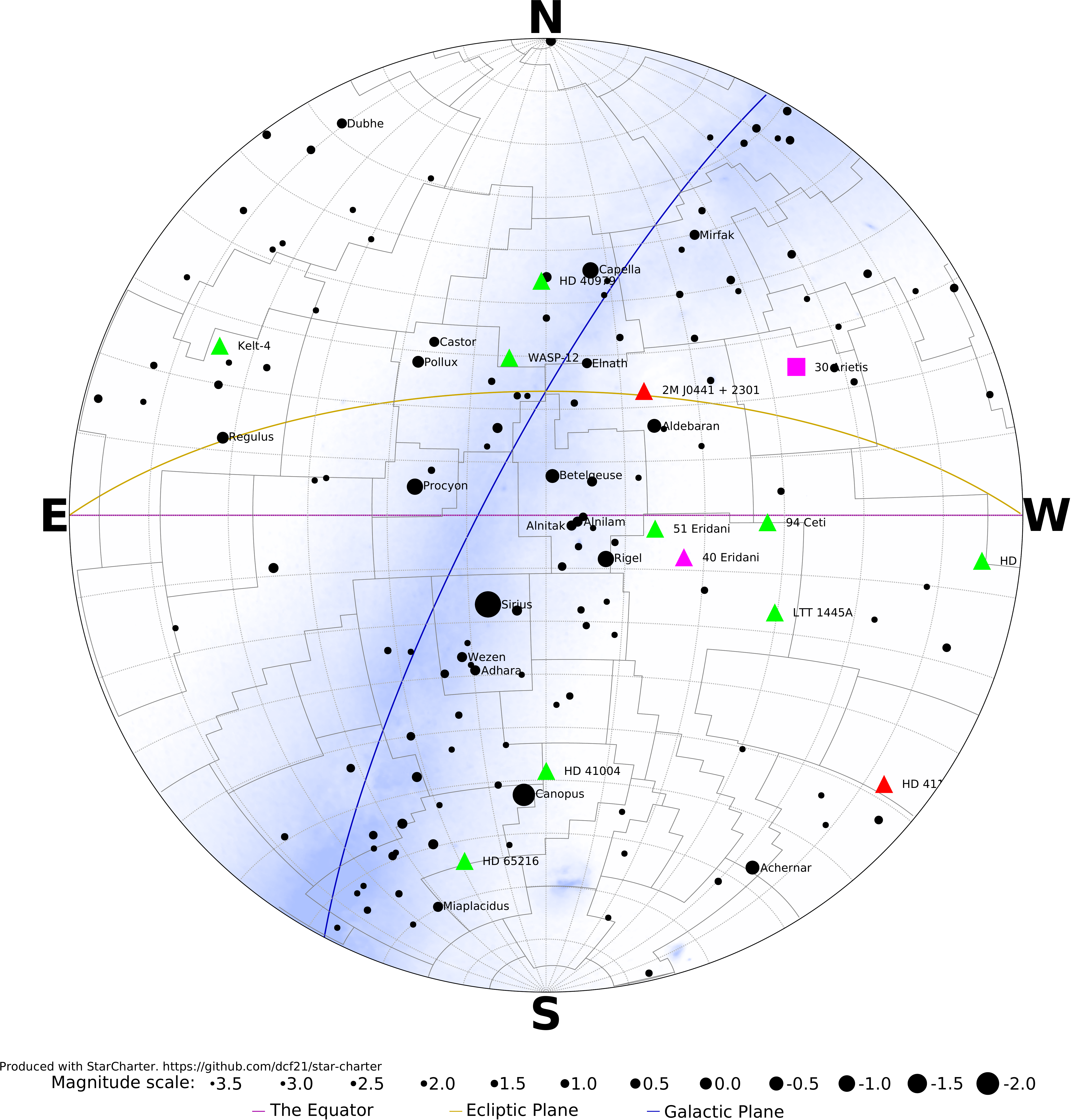}
\caption{
Display of the stellar hemisphere for RA between 0 hrs and 12 hrs.  Stars of different magnitudes
are depicted with the number of stars restricted to about 150.  Additionally, we show the position of
the various triple (triangles) and quadruple stellar systems (squares) with planets.
See Table~4 for information on the color code for the triple stellar systems.
\label{fig:2}}
\end{center}
\end{figure}

\newpage

\begin{figure}[htb]
\begin{center}
\includegraphics[scale=0.35]{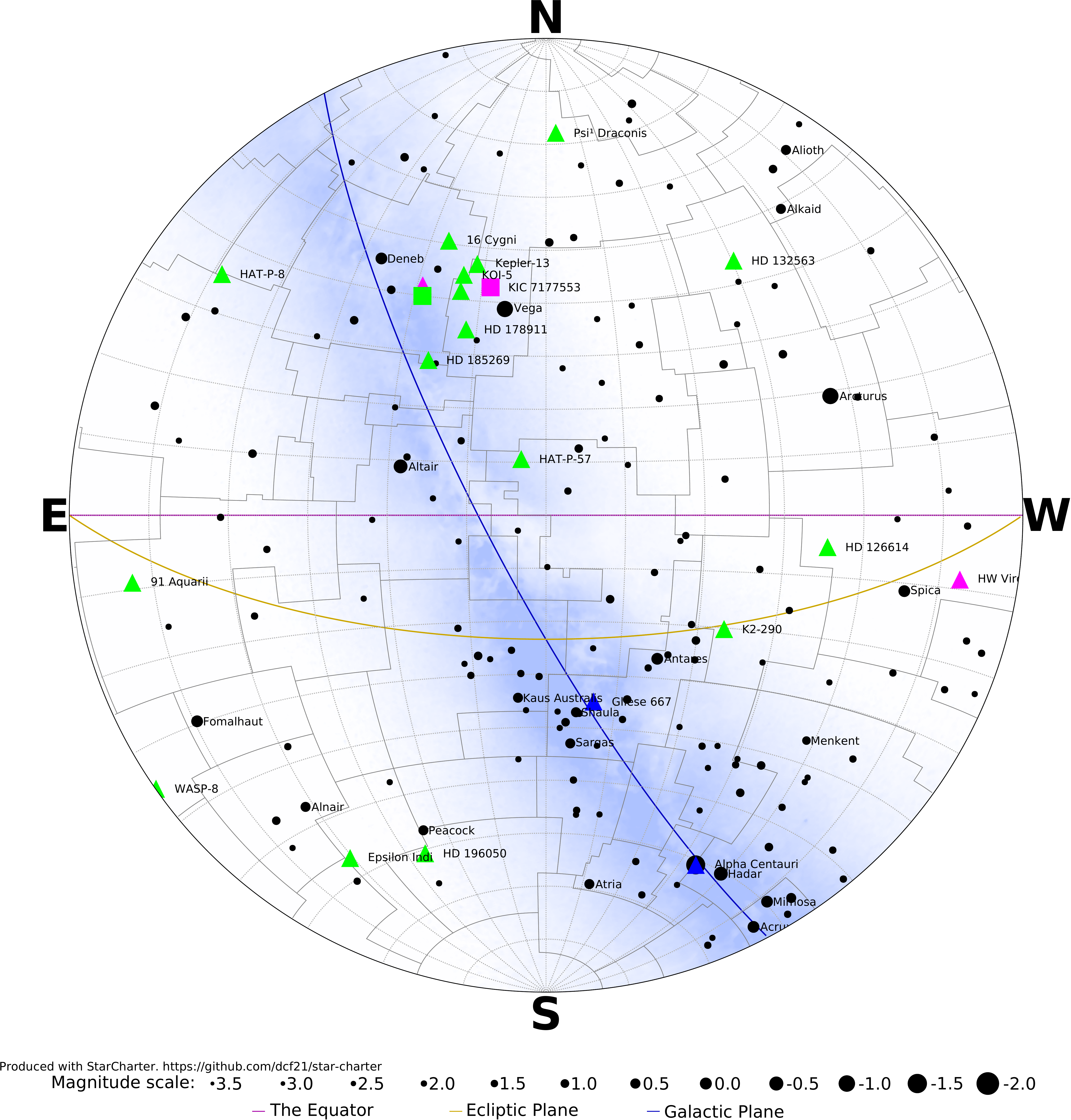}
\caption{
Same as Fig.~2, but for RA between 12 hrs and 24 hrs.  Some of the system names for objects
at or near the Summer Triangle have been omitted due to overcrowding.
\label{fig:3}}
\end{center}
\end{figure}

\newpage

\begin{figure}[htb]
\begin{center}
\includegraphics[scale=1.0]{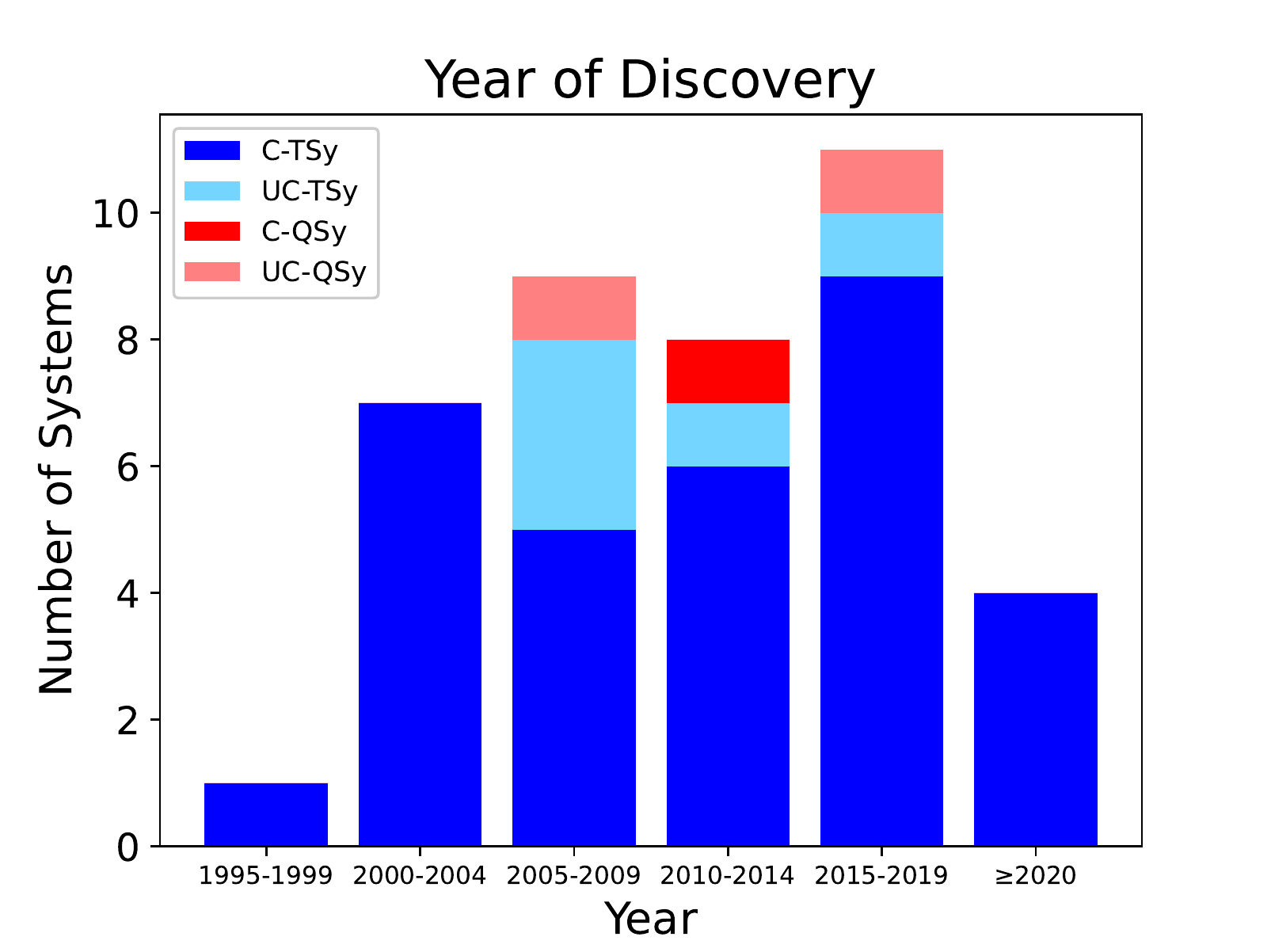}
\caption{
Year of discovery for the planet-hosting triple and quadruple star systems, as gauged
by the year of publication pertaining to the planet(s); see main text for references.
C-TSy and UC-TSy denote certain and uncertain planet-hosting triple star systems, respectively, whereas
C-QSy and UC-QSy denote certain and uncertain planet-hosting quadruple star systems, respectively.
\label{fig:4}}
\end{center}
\end{figure}

\newpage

\begin{figure}[htb]
\begin{center}
\includegraphics[scale=1.0]{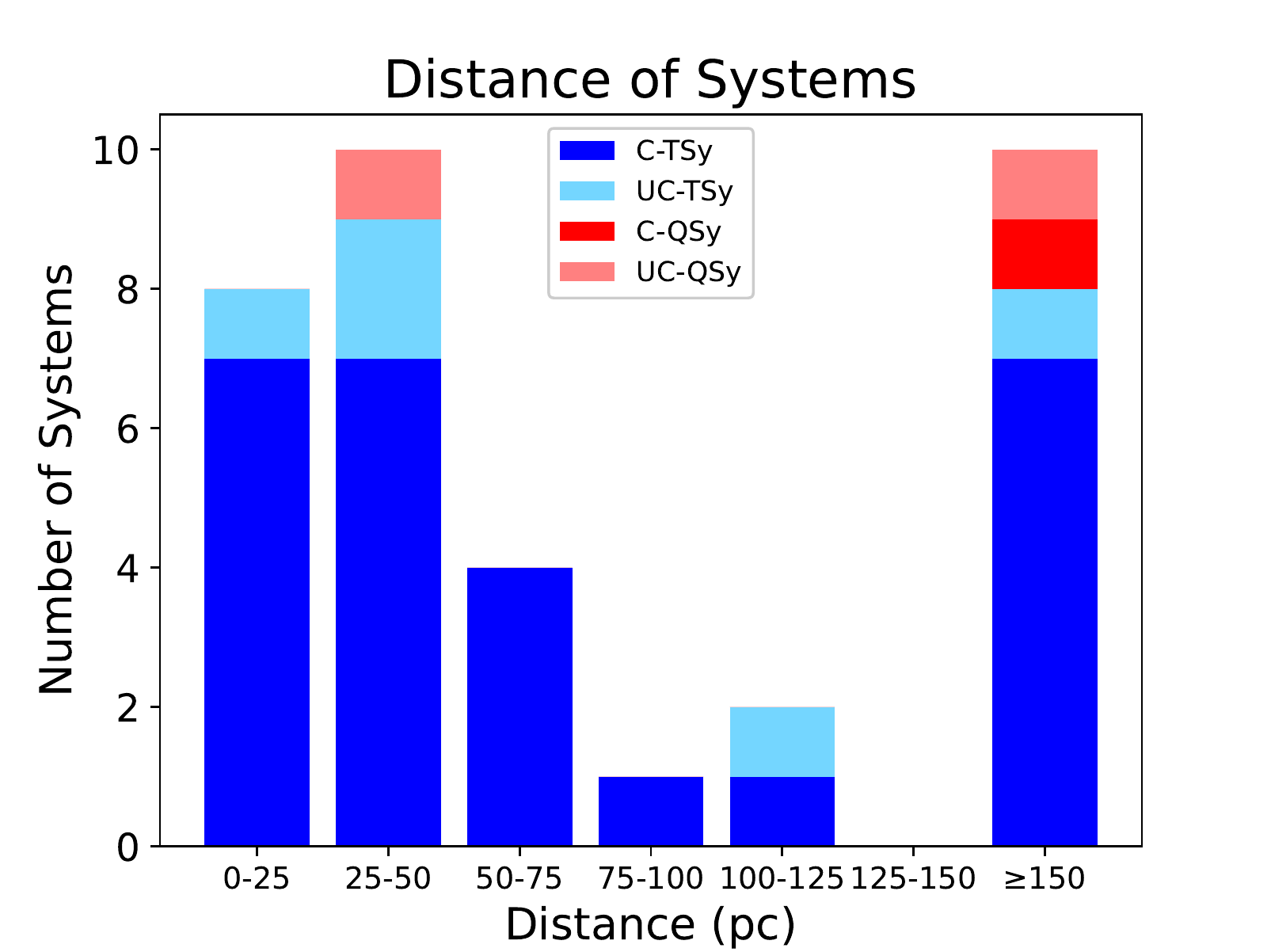}
\caption{
Distance for the various planet-hosting star systems; see main text for references.
See Figure~4 for the definitions of the acronyms.
\label{fig:5}}
\end{center}
\end{figure}

\newpage

\begin{figure}[htb]
\begin{center}
\includegraphics[scale=1.0]{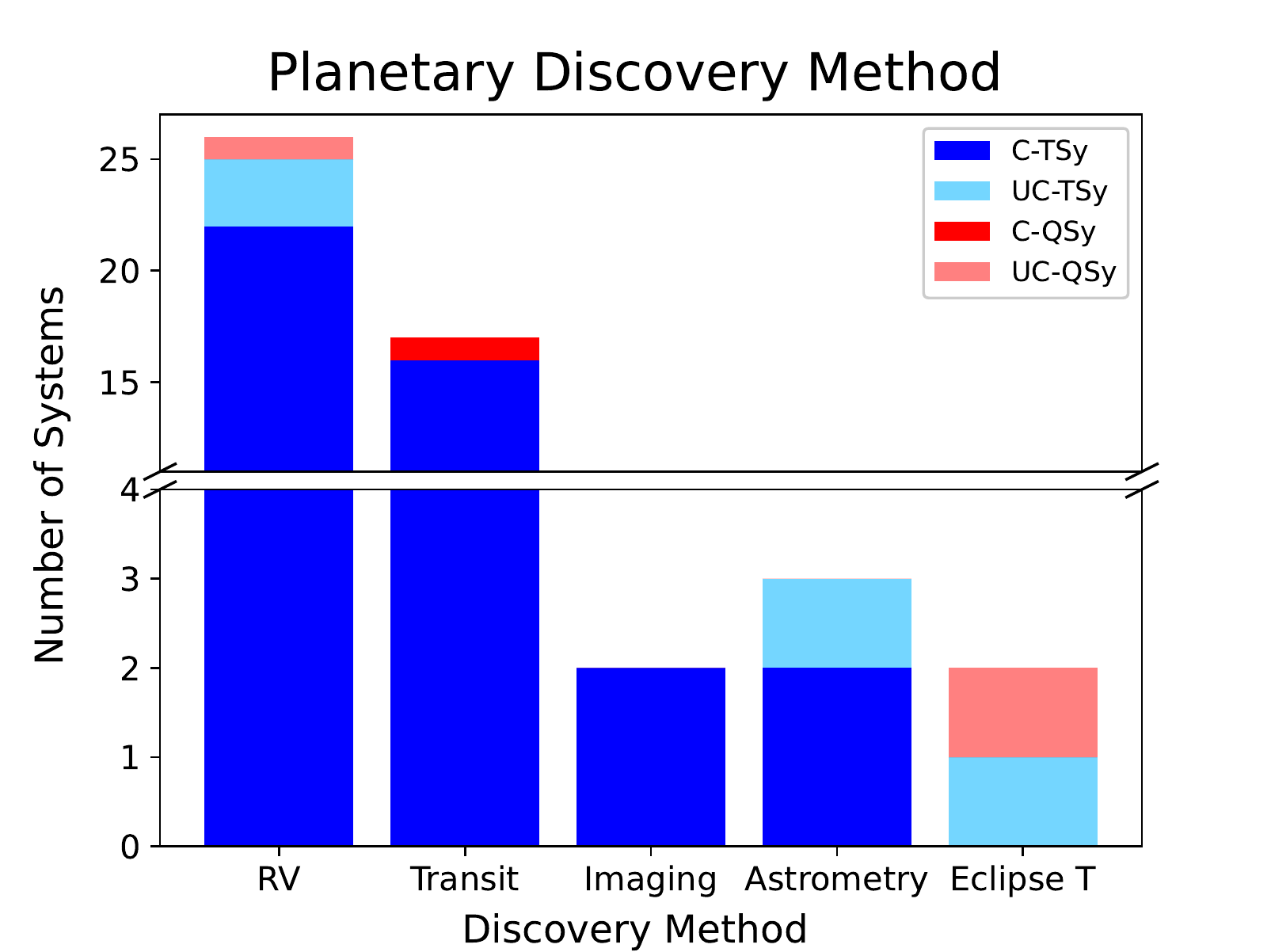}
\caption{
Discovery method for the various exoplanets; see main text for references.
See Figure~4 for the definitions of the acronyms.
\label{fig:6}}
\end{center}
\end{figure}

\newpage

\begin{figure}[htb]
\begin{center}
\includegraphics[scale=1.0]{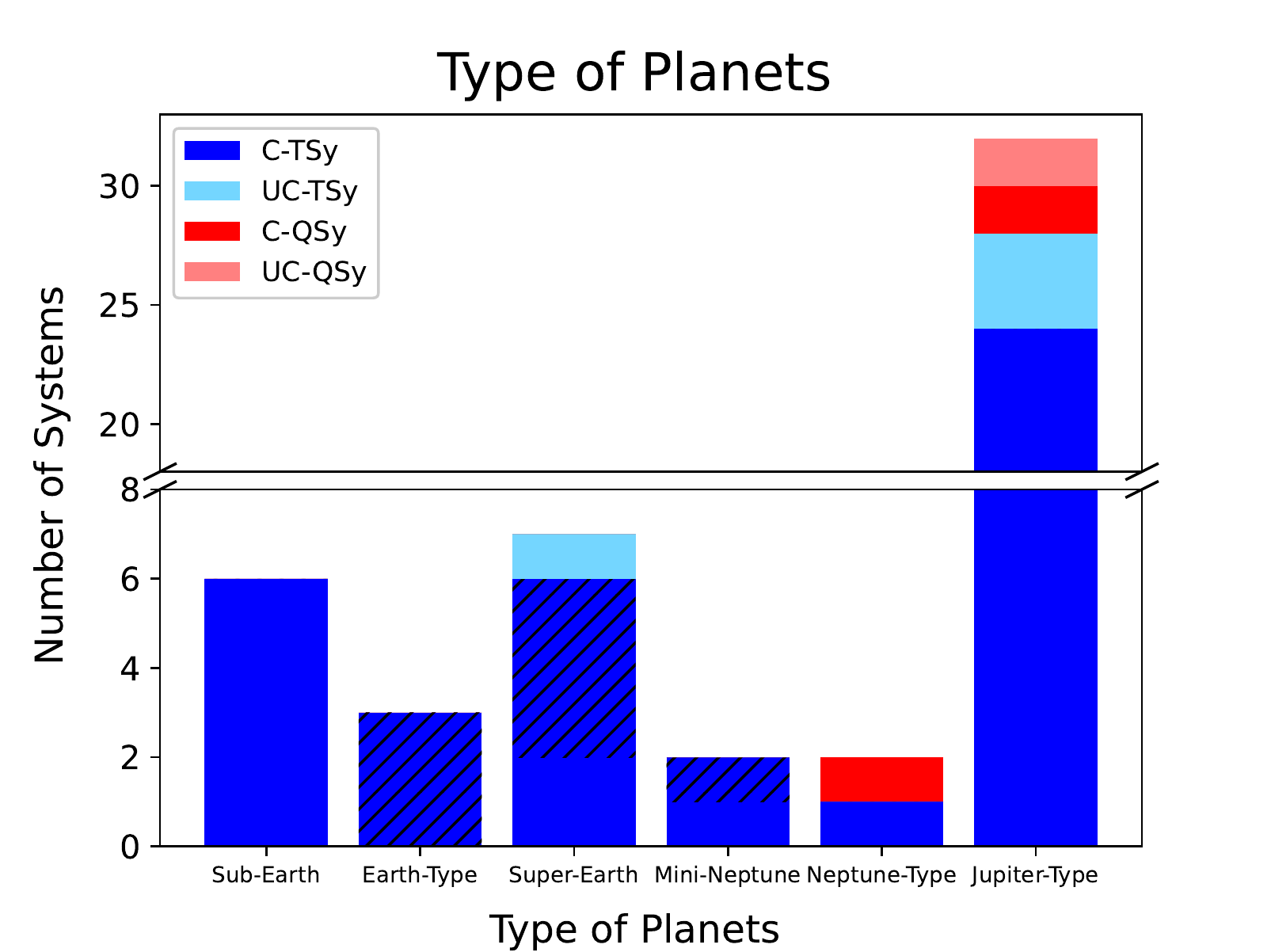}
\caption{
Summary of planetary types; see main text for references.
For some planets, ambiguity exists regarding the planetary types.
In this case, double listing has been chosen with the areas hatched.
See Figure~4 for the definitions of the acronyms.
\label{fig:7}}
\end{center}
\end{figure}

\newpage

\begin{figure}[htb]
\begin{center}
\includegraphics[scale=0.65]{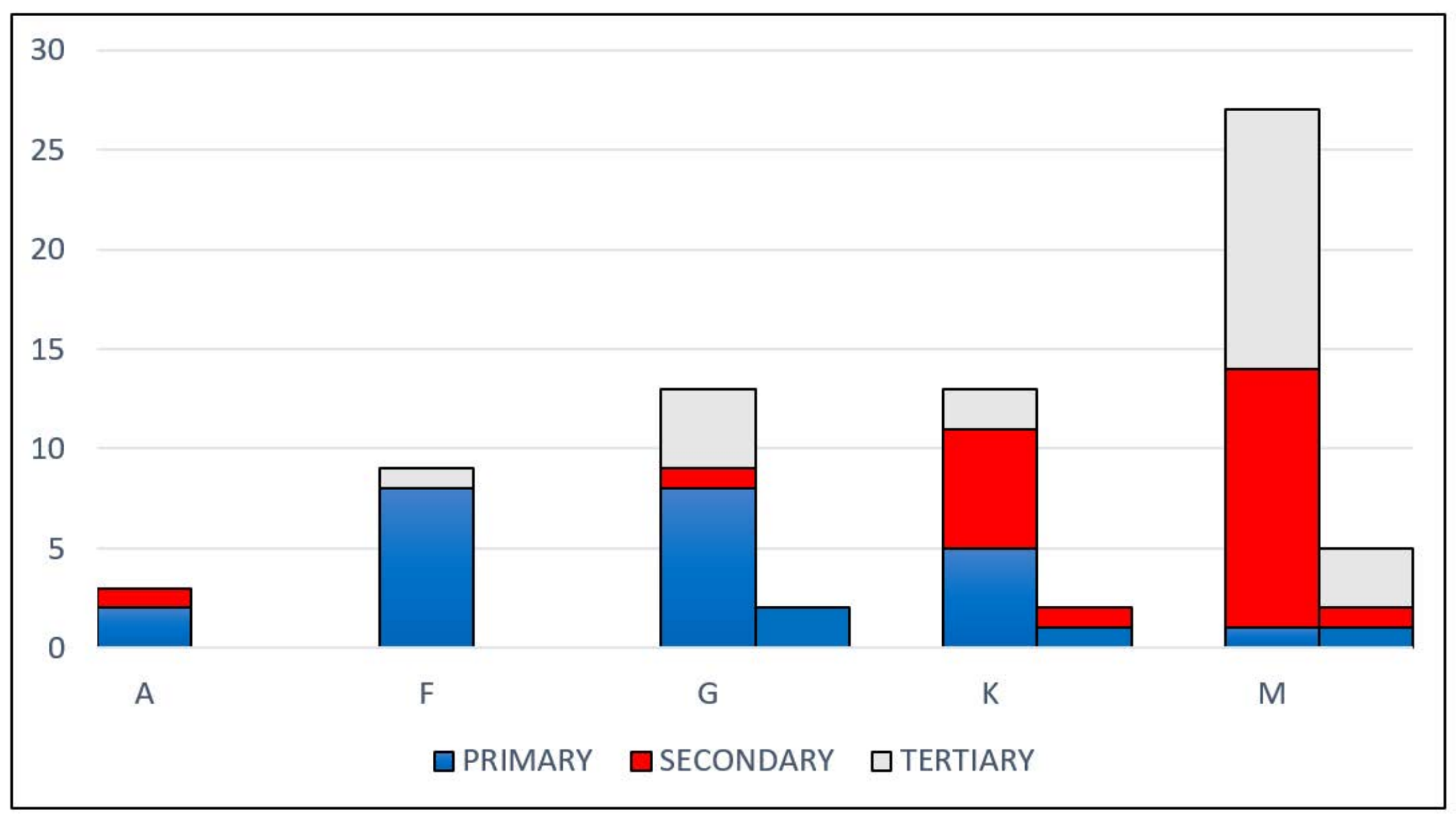}
\caption{
Number count of stellar spectral types for primary, secondary, and tertiary stars
pertaining to planet-hosting triple star systems.  The left bars refer to confirmed systems,
whereas the right bars refer to unconfirmed systems.  There is also a small number
of objects of spectral type L and T not depicted here.
\label{fig:8}}
\end{center}
\end{figure}

\newpage

\begin{figure}[htb]
\begin{center}
\includegraphics[scale=1.0]{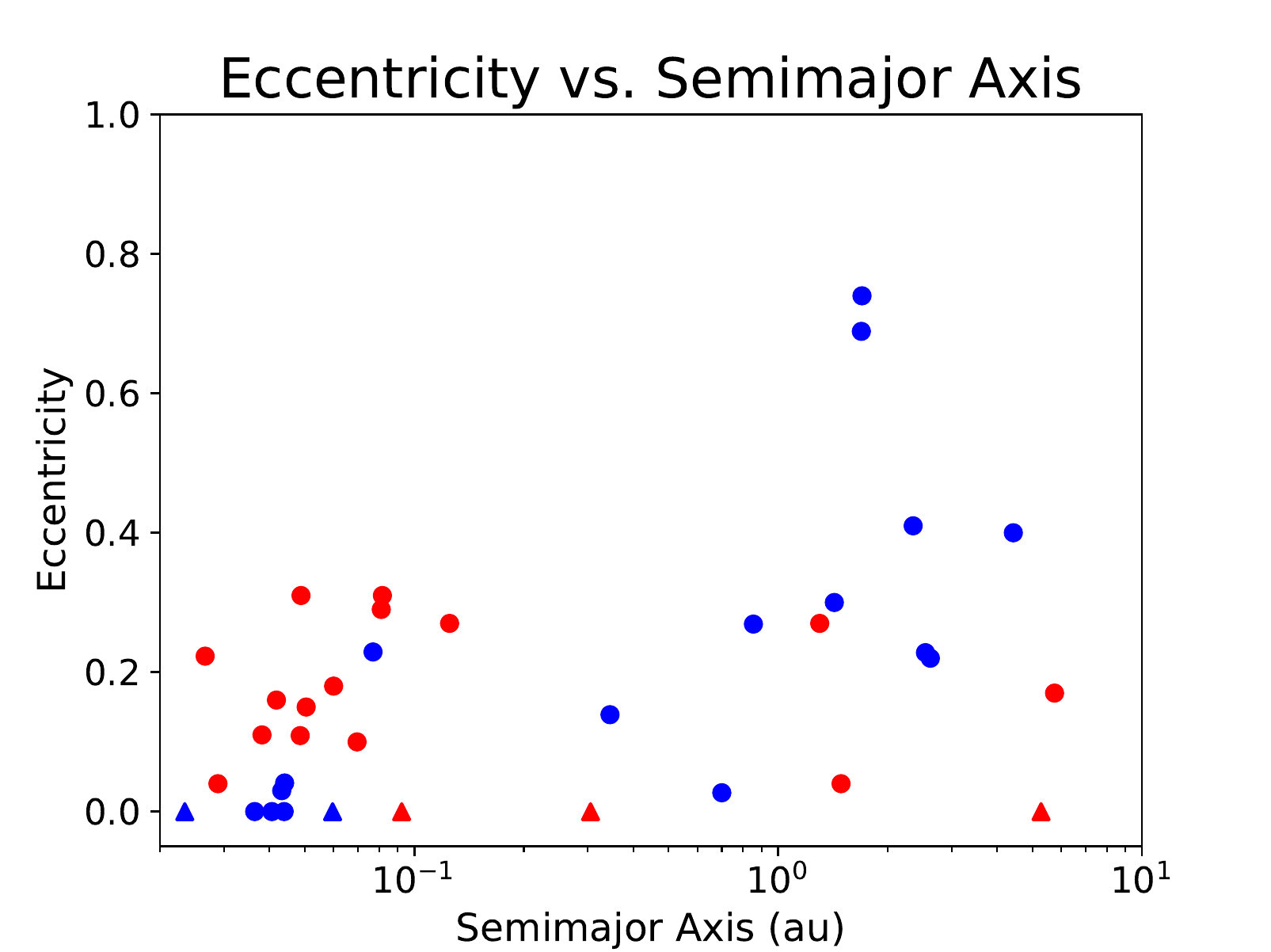}
\caption{
Display of the planetary eccentricity as a function of the semimajor axis
pertaining to confirmed planet-hosting triple star systems.
Planets of single and multiple planetary systems are depicted by blue and red data points, respectively.
If the eccentricity is assumed as zero (instead of measured), a triangle is used.
\label{fig:9}}
\end{center}
\end{figure}

\newpage

\begin{figure}[htb]
\begin{center}
\includegraphics[scale=1.0]{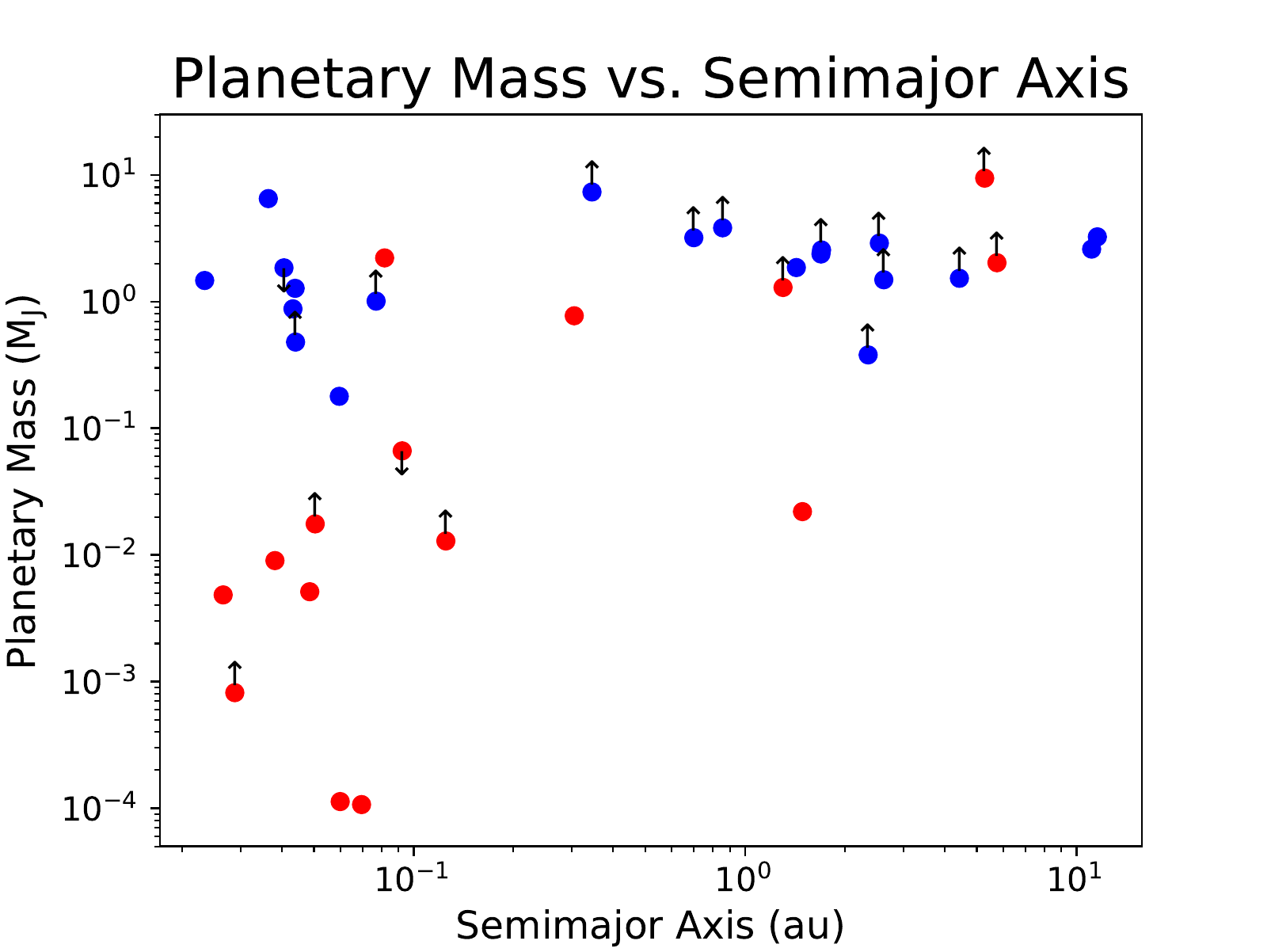}
\caption{
Display of the planet's mass as a function of the semimajor axis
pertaining to confirmed planet-hosting triple star systems.
Planets of single and multiple planetary systems are depicted by blue and red data points, respectively.
An arrow is used when only the planet's minimum or maximum mass is known.
\label{fig:10}}
\end{center}
\end{figure}

\newpage

\begin{figure}[htb]
\begin{center}
\includegraphics[scale=1.0]{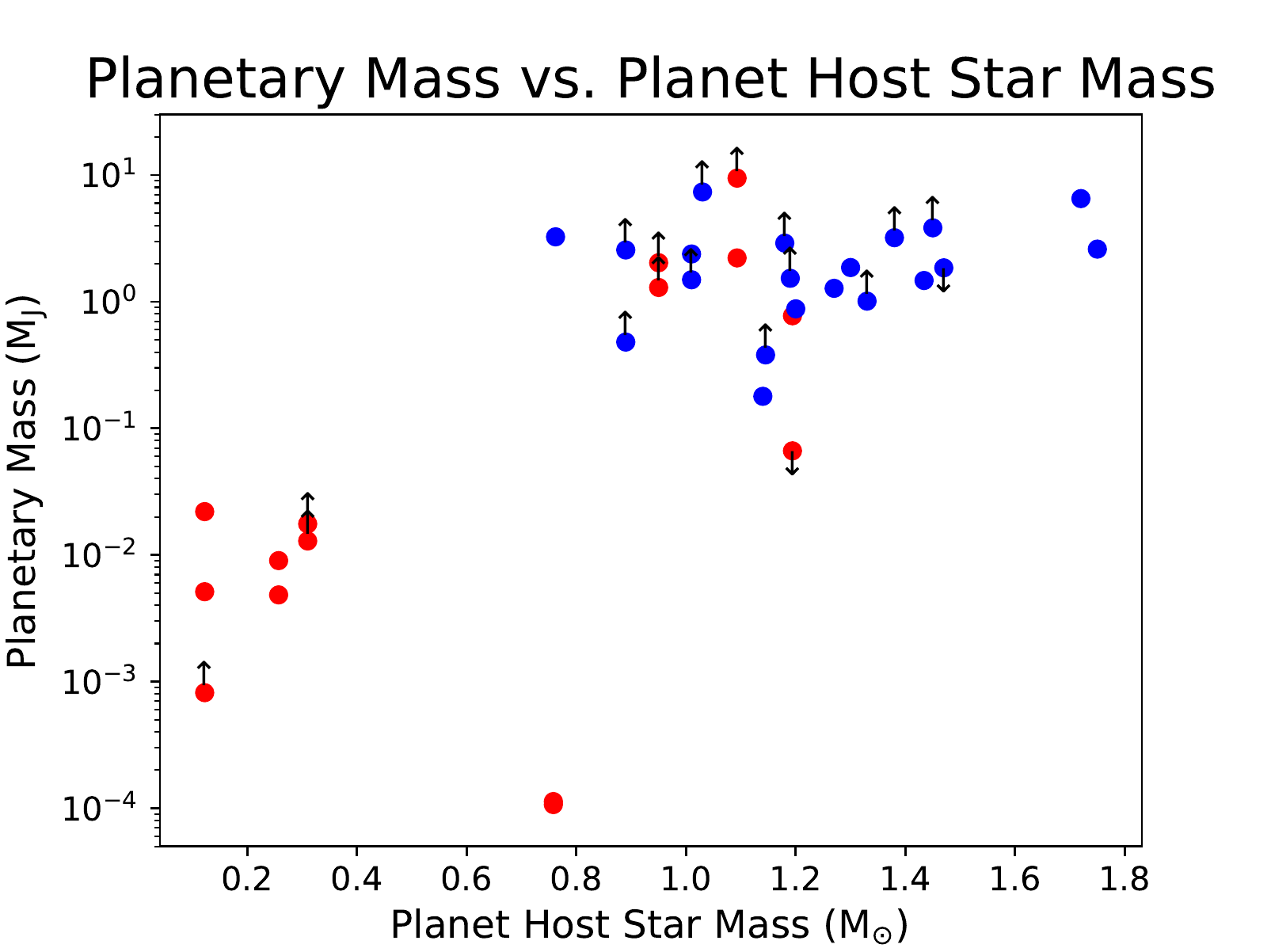}
\caption{
Display of the planet's mass as a function of the host's stellar mass
pertaining to confirmed planet-hosting triple star systems.
Planets of single and multiple planetary systems are depicted by blue and red data points, respectively.
An arrow is used when only the planet's minimum or maximum mass is known.
\label{fig:11}}
\end{center}
\end{figure}

\newpage





\thispagestyle{empty} 
\tabletypesize{\scriptsize}
\begin{deluxetable}{lcclcl}
\tablenum{1}
\tablecaption{Triple and Quadruple Systems, Overview}
\tablewidth{0pt}
\tablehead{
System Name & Type & Planets & Constellation & Distance & Discovery  \\
...         & ...  & ...     & ...           & (pc)     &
}
\startdata
16~Cygni           & 3 & 1     & Cygnus           &   21.1  & COC-97                 \\
2M~J0441~+~2301    & 3 & 0 (1) & Taurus           &  120    & TOD-10                 \\
30~Arietis         & 4 & 0 (1) & Aries            &   44.7  & GUE-09                 \\
40~Eridani         & 3 & 0 (1) & Eridanus         &    5.04 & MA-18                  \\
51~Eridani         & 3 & 1     & Eridanus         &  29.8   & MAC-15                 \\
91~Aquarii         & 3 & 1     & Aquarius         &  44.0   & MIT-03                 \\
94~Ceti            & 3 & 1     & Cetus            &  22.5   & MAY-04                 \\
Alpha~Centauri     & 3 & 3 (1) & Centaurus        &   1.302 & ANG-16, DAM-20, FAR-22 \\
Epsilon~Indi       & 3 & 1     & Indus            &   6.64  & FEN-19                 \\
Gliese~667         & 3 & 2 (1) & Scorpius         &   7.24  & ANG-12                 \\
HAT-P-8            & 3 & 1     & Pegasus          &  212    & LAT-09                 \\
HAT-P-57           & 3 & 1     & Aquila           &  280    & HAR-15                 \\
HD~126614          & 3 & 1     & Virgo            &   73.1  & HOW-10                 \\
HD~132563          & 3 & 1     & Bo{\"o}tes       &  105    & DES-11                 \\
HD~178911          & 3 & 1     & Lyra             &   41.0  & ZUC-02                 \\
HD~185269          & 3 & 1     & Cygnus           &   52.0  & MOU-06                 \\
HD~188753          & 3 & 0 (1) & Cygnus           &   48.1  & KON-05                 \\
HD~196050          & 3 & 1     & Pavo             &   50.7  & JON-02                 \\
HD~2638 / HD~2567  & 3 & 1     & Cetus            &   55.0  & MOU-05                 \\
HD~40979           & 3 & 1     & Auriga           &   34.1  & FIS-03                 \\
HD~41004           & 3 & 1     & Pictor           &   41.6  & ZUC-04                 \\
HD~4113            & 3 & 1     & Sculptor         &   41.9  & TAM-08                 \\
HD~65216           & 3 & 2     & Carina           &   35.1  & MAY-04, WIT-19         \\
HW~Virginis        & 3 & 0 (1) & Virgo            &  172    & LEE-09                 \\
K2-290             & 3 & 2     & Libra            &  273    & HJO-19                 \\
Kelt-4             & 3 & 1     & Leo              &  218    & EAS-16                 \\
Kepler-13          & 3 & 1     & Lyra             &  519    & SHP-11                 \\
Kepler-64          & 4 & 1     & Cygnus           & 1033    & SCH-13                 \\
Kepler-444         & 3 & 5     & Lyra             &   36.4  & CAM-15                 \\
KIC~7177553        & 4 & 0 (1) & Lyra             &  406    & LEH-16                 \\
KOI-5 = TOI-1241   & 3 & 1     & Cygnus           &  547    & CIA-21                 \\
LTT~1445           & 3 & 2     & Eridanus         &    6.87 & WIN-19, WIN-22         \\
Psi$^1$~Draconis   & 3 & 1     & Draco            &   22.7  & END-16                 \\
WASP-8             & 3 & 2     & Sculptor         &   90.0  & QUE-10, KNU-14         \\
WASP-12            & 3 & 1     & Auriga           &  427    & HEB-09                 \\
\enddata
\tablecomments{
See main text for information on the distance measurements, including references.  Values
of higher precision are available for most systems, as well as uncertainty information.
The number of unconfirmed planets is given in parenthesis (see Table~5 for details).
}
\label{table01}
\end{deluxetable}


\clearpage

\tabletypesize{\scriptsize}
\begin{deluxetable}{lcc}
\tablenum{2}
\tablecaption{System Coordinates}
\tablecolumns{3}
\tablewidth{0pt}
\tablehead{
\colhead{System} &
\colhead{Right Ascension} &
\colhead{Declination}\\
\colhead{...} &
\colhead{($^{\circ}$)} &
\colhead{($^{\circ}$)}
}
\startdata
16~Cygni          & 295.454540     &     $+$50.525440  \\
2M~J0441~+~2301   & {\p0}70.437066 &     $+$23.030947  \\
30~Arietis        & {\p0}39.252181 &     $+$24.647219  \\        
40~Eridani        & {\p0}63.817999 & {\p0}$-$7.652872  \\
51~Eridani        & {\p0}69.400551 & {\p0}$-$2.473550  \\           
91~Aquarii        &  348.972897    & {\p0}$-$9.087734  \\         
94~Ceti           & {\p0}48.193487 & {\p0}$-$1.196097  \\ 
Alpha~Centauri    &  219.873833    &     $-$60.832222  \\  
Epsilon~Indi      &  330.840226    &     $-$56.785983  \\          
Gliese~667        &  259.738187    &     $-$34.989762  \\      
HAT-P-8           &  343.041098    &     $+$35.447113  \\            
HAT-P-57          &  274.743438    &     $+$10.597258  \\         
HD~126614         &  216.701165    & {\p0}$-$5.177781  \\          
HD~132563         &  224.589654    &     $+$44.043146  \\         
HD~178911         &  287.268133    &     $+$34.600345  \\          
HD~185269         &  294.298921    &     $+$28.499862  \\        
HD~188753         &  298.743216    &     $+$41.871536  \\  
HD~196050         &  309.465458    &     $-$60.634485  \\        
HD~2638 / HD~2567 & {\pd0}7.317127 & {\p0}$-$5.910490  \\            
HD~40979          & {\p0}91.124762 &     $+$44.260444  \\          
HD~41004          & {\p0}89.956864 &     $-$48.239668  \\        
HD~4113           & {\p0}10.802482 &     $-$37.982633  \\  
HD~65216          &  118.422164    &     $-$63.647320  \\
HW~Virginis       &  191.084328    & {\p0}$-$8.671347  \\  
K2-290            &  234.857760    &     $-$20.198825  \\       
Kelt-4            &  157.062546    &     $+$25.573187  \\      
Kepler-13         &  286.971271    &     $+$46.868335  \\       
Kepler-64         &  298.215071    &     $+$39.955103  \\      
Kepler-444        &  289.752288    &     $+$41.634602  \\     
KIC~7177553       &  283.012120    &     $+$42.721259  \\
KOI-5 = TOI-1241  &  289.739713    &     $+$44.647394  \\
LTT~1445          & {\p0}45.462500 &     $-$16.593364  \\
Psi$^1$~Draconis  &  265.484625    &     $+$72.149500  \\  
WASP-8            &  359.900297    &     $-$35.031368  \\  
WASP-12           & {\p0}97.636653 &     $+$29.672296  \\  
\enddata
\tablecomments{
Values are for J2000.  The right ascension (RA) is given in units of degrees ($^{\circ}$) rather than hours (h).
}
\label{table02}
\end{deluxetable}


\thispagestyle{empty} 
\tabletypesize{\scriptsize}
\begin{deluxetable}{lccll}
\tablenum{3}
\tablecaption{Method of Planet Discovery and Facility}
\tablewidth{0pt}
\tablehead{
System Name & Type & Planets & Method & Facility
}
\startdata
16~Cygni           & 3 & 1     & RV                & McDonald Obs., Lick Obs.              \\
2M~J0441~+~2301    & 3 & 0 (1) & Astrometry        & HST, Gemini Obs.                      \\
30~Arietis         & 4 & 0 (1) & RV                & Karl Schwarzschild Obs.               \\
40~Eridani         & 3 & 0 (1) & RV                & Keck Obs., Dharma Planet Imager       \\
51~Eridani         & 3 & 1     & Imaging           & Gemini Obs.                           \\ 
91~Aquarii         & 3 & 1     & RV                & Lick Obs.                             \\
94~Ceti            & 3 & 1     & RV                & La Silla Obs.                         \\
Alpha~Centauri$^a$ & 3 & 3 (1) & ...               & ...                                   \\ 
Epsilon~Indi       & 3 & 1     & RV, Astrometry    & La Silla Obs., {\it Gaia}             \\ 
Gliese~667         & 3 & 2 (1) & RV                & La Silla Obs.                         \\ 
HAT-P-8            & 3 & 1     & Transit           & Keck Obs.                             \\ 
HAT-P-57           & 3 & 1     & Transit           & Keck Obs.                             \\ 
HD~126614          & 3 & 1     & RV                & Keck Obs.                             \\ 
HD~132563          & 3 & 1     & RV                & Galileo Nat'l Telescope               \\ 
HD~178911          & 3 & 1     & RV                & Keck Obs., Obs. de Haute-Provence     \\ 
HD~185269          & 3 & 1     & RV                & Obs. de Haute-Provence                \\ 
HD~188753          & 3 & 0 (1) & RV                & Keck Obs.                             \\
HD~196050          & 3 & 1     & RV                & La Silla Obs.                         \\ 
HD~2638 / HD~2567  & 3 & 1     & RV                & La Silla Obs.                         \\ 
HD~40979           & 3 & 1     & RV                & Lick Obs., Keck Obs.                  \\ 
HD~41004           & 3 & 1     & RV                & La Silla Obs.                         \\ 
HD~4113            & 3 & 1     & RV                & La Silla Obs.                         \\
HD~65216           & 3 & 2     & RV                & La Silla Obs.                         \\ 
HW~Virginis        & 3 & 0 (1) & Eclipse Timing    & SOAO, CNUO                            \\
K2-290             & 3 & 2     & Transit           & K2                                    \\ 
Kelt-4             & 3 & 1     & Transit           & SuperWASP$^c$                         \\ 
Kepler-13          & 3 & 1     & Transit           & {\it Kepler}                          \\ 
Kepler-64          & 4 & 1     & Transit           & {\it Kepler}                          \\
Kepler-444         & 3 & 5     & Transit           & {\it Kepler}                          \\
KIC~7177553        & 4 & 0 (1) & Eclipse Timing    & {\it Kepler}, Karl Schwarzschild Obs. \\
KOI-5 = TOI-1241   & 3 & 1     & Transit / Imaging & {\it Kepler}, TESS                    \\ 
LTT~1445           & 3 & 2     & Transit           & TESS$^b$                              \\ 
Psi$^1$~Draconis   & 3 & 1     & RV                & McDonald Obs.                         \\ 
WASP-8             & 3 & 2     & Transit / RV      & SuperWASP$^c$, Keck Obs.              \\ 
WASP-12            & 3 & 1     & Transit           & SuperWASP$^c$                         \\ 
\enddata
\tablecomments{
See Table~1 for further information, including references.
Besides the facilities as indicated, other sites have typically been highly relevant as well,
especially for the establishment and confirmation of the system's multiplicity and the properties
of the system's components. \\
$^a$ The three confirmed planets have been identified through the RV method, except Proxima Centauri~b
where astrometry contributed as well.  Relevant facilities included La Silla, VLT, and HST. \\
$^b$ For LTT~1445Ac, additional data from five spectrographs have been used to established the planet's existence. \\
$^c$ SuperWASP consists of two robotic telescopes, located on the island of La Palma and at the site of the
South African Astronomical Observatory.
}
\label{table03}
\end{deluxetable}


\clearpage

\thispagestyle{empty} 
\begin{deluxetable}{lcl}
\tablenum{4}
\tablecaption{Case Legend for the Star--Planet Systems}
\tablewidth{0pt}
\tablehead{
Type           & Color            &  Definition
}
\startdata
\noalign{\smallskip}
\hline
\noalign{\smallskip}
OK Case        &    Green     & No noted controversy about the existence of the        \\
               &              & ~~~stellar components or the planet(s).                \\
Case~1         &    Blue      & The stellar components are confirmed and at least      \\
               &              & ~~~one exoplanet is confirmed; however, there may be   \\
               &              & ~~~another unconfirmed exoplanet or stellar component. \\
Case~2         &    Red       & There is an unconfirmed stellar component, with        \\
               &              & ~~~one or more confirmed exoplanets.                   \\
Case~3         &    Purple    & The stellar components are confirmed; however,         \\
               &              & ~~~the exoplanet is unconfirmed.                       \\
\noalign{\smallskip}
\hline
\noalign{\smallskip}
Confirmed System      &    ...       & OK Case and Case~1                              \\
Unconfirmed System    &    ...       & Case~2 and Case~3                               \\
\noalign{\smallskip}
\hline
\noalign{\smallskip}
\enddata
\tablecomments{
The denoted color scheme has been used in Figs.~2 and 3.
}
\label{table04}
\end{deluxetable}


\clearpage

%
\thispagestyle{empty}
\tabletypesize{\scriptsize}
\begin{deluxetable}{lcccc}
\tablenum{5}
\tablecaption{Triple and Quadruple System Classification}
\tablewidth{0pt}
\tablehead{
System & Type                 & Uncertainty & \multicolumn{2}{c}{Orbit Type}  \\
...    & ...                  & ...         & Confirmed & Unconfirmed
}
\startdata
16~Cygni              &   3   & ...     &  S2             &  ...            \\
2M~J0441~+~2301$^a$   &   3   & Case~2  &  ...            &  S2             \\
30~Arietis$^b$        &   4   & Case~3  &  ...            &  S3             \\
40~Eridani$^c$        &   3   & Case~3  &  ...            &  S1             \\
51~Eridani            &   3   & ...     &  S1             &  ...            \\
91~Aquarii            &   3   & ...     &  S1             &  ...            \\
94~Ceti               &   3   & ...     &  S1             &  ...            \\
Alpha~Centauri$^d$    &   3   & Case~1  &  S3 (3$\times$) &  S1, S2         \\
Epsilon~Indi          &   3   & ...     &  S1             &  ...            \\
Gliese~667$^e$        &   3   & Case~1  &  S3 (2$\times$) &  S3             \\
HAT-P-8               &   3   & ...     &  S1             &  ...            \\
HAT-P-57              &   3   & ...     &  S1             &  ...            \\
HD~126614             &   3   & ...     &  S1             &  ...            \\
HD~132563             &   3   & ...     &  S3             &  ...            \\
HD~178911             &   3   & ...     &  S3             &  ...            \\
HD~185269             &   3   & ...     &  S1             &  ...            \\
HD~188753$^f$         &   3   & Case~3  &  ...            &  S1             \\
HD~196050             &   3   & ...     &  S1             &  ...            \\
HD~2638 / HD~2567     &   3   & ...     &  S2             &  ...            \\
HD~40979              &   3   & ...     &  S1             &  ...            \\
HD~41004              &   3   & ...     &  S1             &  ...            \\
HD~4113$^g$           &   3   & Case~2  &  ...            &  S1             \\
HD~65216              &   3   & ...     &  S1 (2$\times$) &  ...            \\
HW~Virginis$^h$       &   3   & Case~3  &  ...            &  P12            \\
K2-290                &   3   & ...     &  S1 (2$\times$) &  ...            \\
Kelt-4                &   3   & ...     &  S1             &  ...            \\
Kepler-13             &   3   & ...     &  S1             &  ...            \\
Kepler-64             &   4   & ...     &  P12            &  ...            \\
Kepler-444            &   3   & ...     &  S1 (5$\times$) &  ...            \\
KIC~7177553$^i$       &   4   & Case~3  &  ...            &  P12            \\
KOI-5 = TOI-1241      &   3   & ...     &  S1             &  ...            \\
LTT~1445              &   3   & ...     &  S1 (2$\times$) &  ...            \\
Psi$^1$~Draconis      &   3   & ...     &  S2             &  ...            \\
WASP-8                &   3   & ...     &  S1 (2$\times$) &  ...            \\
WASP-12               &   3   & ...     &  S1             &  ...            \\
\enddata
\tablecomments{See Table~4 for information on the definition of Case~1 to 3. \\
$^a$ Unclear if one of the system components is a low-mass brown dwarf or a giant planet. \\
$^b$ The suspected planet may in fact be a brown dwarf or red dwarf.  In this case, the system would thus be a quintuple system
without a planet instead of a planet-hosting quadruple system (see Sect.~3.2). \\
$^c$ There is evidence that the previously reported planet is a false positive. \\
$^d$ There are claims of additional planets regarding all three stellar components. \\
$^e$ Two planets are confirmed; however, there are claims of at least one additional planet orbiting the same component. \\
$^f$ Follow-up observations did not confirm the planet's existence, which may however be due to the limited precession of that measurement. \\
$^g$ Some members of the system are not fully established; hence, HD~4113 may be a planet-hosting binary system. \\
$^h$ The system is potentially dynamically unstable.  Apparently, it consists of two stars, a Jupiter-type planet, and a brown dwarf;
thus, a triple star system. \\
$^i$ More transit data are needed to ultimately confirm the planet's existence.
}
\label{table05}
\end{deluxetable}


\clearpage

\thispagestyle{empty} 
\tabletypesize{\scriptsize}
\begin{deluxetable}{lccccccll}
\tablenum{6}
\tablecaption{Triple System Configurations}
\tablewidth{0pt}
\tablehead{
System Name  & Status  & Star~1 & Star~2 & Star~3 & ${a_{\rm bin}}^{(1)}$ & ${a_{\rm bin}}^{(2)}$ & Configuration & Reference \\
...          & ...     & ...    & ...    & ...    & (au)                  & (au)                  & ...           & ...
}
\startdata
 16~Cygni              &      & G1.5~V    & M          & G2.5~V     &     73   &   860   & AC-B, 2-1    & RAG-06          \\
 2M~J0441~+~2301       &  UC  & M8.5~V    & ...        & M3.5~V     &     ...  &   ...   & A-BC, 1-2    & TOD-10          \\        
 40~Eridani            &  UC  & K0~V      & DA2.9      & M4.5~V     &     35   &   400   & A-BC, 1-2    & MAS-17          \\
 51~Eridani            &      & F0~V      & M          & M          &     9.8  &  2000   & A-BC, 1-2    & FEI-06, MON-15  \\   
 91~Aquarii            &      & K1~III    & K8~V       & K          &    21.5  &  2248   & A-BC, 1-2    & RAG-06, ZIR-07  \\    
 94~Ceti               &      & F8~V      & M0~V       & M3~V       &     0.98 &   220   & A-BC, 1-2    & ROB-11, ROE-12  \\
 Alpha~Centauri        &      & G2~V      & K1~V       & M5.5~Ve    &    17.5  &  8200   & AB-C, 2-1    & AKE-21          \\    
 Epsilon~Indi          &      & K5~V      & T1-1.5     & T6         &     2.6  &  1459   & A-BaBb, 1-2  & SCH-03, DIE-18  \\ 
 Gliese~667            &      & K3~V      & K5~V       & M1.5~V     &     1.82 &   300   & AB-C, 2-1    & FER-14          \\ 
 HAT-P-8               &      & F5~V      & M5~V       & M6~V       &    15    &   ...   & A-BC, 1-2    & BEC-14          \\
 HAT-P-57              &      & A8~V      & ...        & ...        &    68    &   800   & A-BC, 1-2    & HAR-15          \\   
 HD~126614             &      & K0~V      & M          & M5.5~V     &    36    &  3040   & AB-C, 2-1    & DEA-14, LOD-14  \\
 HD~132563             &      & F8~V      & M          & G0~V       &    40    &   400   & AaAb-B, 2-1  & DES-11          \\
 HD~178911             &      & G1~V      & K1~V       & G5~V       &     4.9  &   789   & AaAb-B, 2-1  & RAG-06          \\
 HD~185269             &      & G0~IV     & ...        & ...        &     5    &   ...   & A-BaBb, 1-2  & GIN-16          \\
 HD~188753             &  UC  & G8~V      & K0~V       & ...        &     0.67 &    12.3 & A-BaBb, 1-2  & POR-05          \\
 HD~196050             &      & G3~V      & M1.5-4.5~V & M2.5-4.5~V &    28    &  7511   & A-BaBb, 1-2  & MAS-01, MUG-05  \\ 
 HD~2638 / HD~2567     &      & G0~V      & G8~V       & M1~V       &    ...   &    25.5 & A-BC, 1-2    & ROB-15          \\
 HD~40979              &      & F8~V      & K3~V       & M3~V       &   129    &  6416   & A-BC, 1-2    & MUG-07          \\
 HD~41004              &      & K1~V      & M2~V       & ...        &  0.017   &    21   & A-BBb, 1-2   & SAN-02          \\  
 HD~4113               &  UC  & G5~V      & T9         & M0-1~V     &    23    &  2000   & AC-B, 2-1    & CHE-18          \\
 HD~65216              &      & G5~V      & M7-8~V     & L2-3       &     6    &   253   & A-BC, 1-2    & MUG-07          \\
 HW~Virginis           &  UC  & sdB       & M6-7~V     & ...        &     0.86 &   ...   & AB-C, 2-1    & LEE-09          \\
 Kelt-4                &      & F         & K          & K          &    10.3  &   328   & A-BC, 1-2    & EAS-16          \\
 Kepler-13             &      & A0~V      & A          & G          &    ...   &   610   & A-BC, 1-2    & SHP-14          \\ 
 Kepler-444            &      & K0~V      & M          & M          &     0.3  &   36.7  & A-BC, 1-2    & DUP-16          \\
 KOI-5 = TOI-1241      &      & G         &  ...       & ...        &    ...   &   ...   & AB-C, 2-1    & ...             \\
 K2-290                &      & F8~V      & M          & M          &   113    &  2467   & AB-C, 2-1    & HJO-19          \\  
 LTT~1445              &      & M2.5~V    & M3~V       & M          &     1.2  &    34   & A-BC, 1-2    & WIN-19          \\
 Psi$^1$~Draconis      &      & F5~IV-V   & ...        & F8~V       &   ...    &   680   & AC-B, 2-1    & GUL-15          \\
 WASP-8                &      & G6~V      & M          & ...        &   ...    &   408   & AB-C, 2-1    & BOH-20          \\
 WASP-12               &      & F         & M3~V       & M3~V       &    21    &   ...   & A-BC, 1-2    & BEC-14          \\
\enddata
\tablecomments{UC indicates that the status of the system is unconfirmed (i.e., Case 2 or 3).
Stars with the luminosity class omitted are almost certainly main-sequence stars.
Configuration information employs the acronyms of the stellar components readily
used in the literature.  Background information and additional references,
including those pertaining to spectral types, can be found in the articles
reporting the respective planet discoveries (see Table~1)
as well as in the main text, especially Sect.~2.3 and 3.1.
As all triple star systems are hierarchic, ${a_{\rm bin}}^{(1)}$ and ${a_{\rm bin}}^{(2)}$
denote the small and large separation distances of the stellar components (which are both often
{\it relatively uncertain} and may also be affected by {\it projection effects}), respectively;
see Sect. 2.2.3 for additional comments.
The references pertain to the system's composition, including
${a_{\rm bin}}^{(1)}$ and ${a_{\rm bin}}^{(2)}$.
}
\label{table06}
\end{deluxetable}


\clearpage

%
\thispagestyle{empty} 
\tabletypesize{\normalsize}
\begin{deluxetable}{lcc}
\tablenum{7}
\tablecaption{Planetary Orbital Classification, Triple Systems}
\tablewidth{0pt}
\tablehead{
Type   & Confirmed &  Unconfirmed
}
\startdata
S1     &  28   &  4    \\
S2     &   3   &  2    \\
S3     &   7   &  1    \\
P12    &   0   &  1    \\
P13    &   0   &  0    \\
P23    &   0   &  0    \\
T123   &   0   &  0    \\
\enddata
\label{table07}
\end{deluxetable}


\clearpage

\thispagestyle{empty} 
\tabletypesize{\scriptsize}
\begin{deluxetable}{lcccccl}
\tablenum{8}
\tablecaption{Triple Systems, Confirmed --- Planet Host Star Data}
\tablewidth{0pt}
\tablehead{
System Name  & Spectral Type & Temperature & Mass        & Radius      & Luminosity  & Reference \\
...          & ...           & (T)         & ($M_\odot$) & ($R_\odot$) & ($L_\odot$) & ...
}
\startdata
16~Cygni             & G2.5~V  &  5751  & 1.01   &  1.12   &  1.25     & TUC-14, MET-15                 \\
51~Eridani           & F0~V    &  7331  & 1.75   &  1.45   &  5.45     & SIM-11, RAJ-17, ROS-20         \\
91~Aquarii           & K1~III  &  4730  & 1.38   & 10.96   & 54.3      & BAI-18                         \\
94~Ceti              & F8~V    &  6055  & 1.30   &  1.90   &  4.02     & KOV-03, FUH-08, BEL-09, BOY-13 \\
Alpha~Centauri       & M5.5~Ve &  2980  & 0.120  &  0.146  &  0.00151  & BES-91, SEG-03, BOY-12, MAN-15, RIB-17 \\
Epsilon~Indi         & K5~V    &  4649  & 0.762  &  0.71   &  0.21     & DEM-09, RAI-20                 \\
Gliese~667           & M1.5~V  &  3700  & 0.31   &  0.42   &  0.0137   & PAS-01, ANG-12                 \\
HAT-P-8              & F       &  6410  & 1.27   &  1.49   &  3.37     & WAN-21                         \\
HAT-P-57             & A8~V    &  7500  & 1.47   &  1.5    &  6.4      & HAR-15                         \\
HD~126614            & K0~V    &  5585  & 1.145  &  1.09   &  1.21     & HOW-10                         \\
HD~132563            & G0~V    &  5985  & 1.01   &  ...    &   ...     & DES-04, DES-11                 \\
HD~178911            & G5~V    &  5642  & 1.03   &  1.05   &  1.00     & BON-16                         \\
HD~185269            & G0~IV   &  5983  & 1.33   &  ...    &   ...     & MOU-06                         \\
HD~196050            & G3~V    &  5920  & 1.18   &  1.46   &  2.21     & CHA-19                         \\
HD~2638 / HD~2567    & G8~V    &  5160  & 0.89   &  0.80   &  0.41     & BON-15                         \\
HD~40979             & F8~V    &  6077  & 1.45   &  1.26   &  1.96     & STA-17                         \\
HD~41004             & K1~V    &  5255  & 0.89   &  1.04   &  0.63     & SOU-18                         \\
HD~65216             & G5~V    &  5718  & 0.95   &  0.86   &  0.72     & BON-15                         \\
K2-290               & F8~V    &  6302  & 1.19   &  1.51   &  3.23     & HJO-19                         \\
Kelt-4               & F       &  6206  & 1.20   &  1.61   &  3.44     & EAS-16                         \\
Kepler-13            & A0~V    &  7650  & 1.72   &  1.71   &  8.99     & SHP-14                         \\
Kepler-444           & K0~V    &  5046  & 0.758  &  0.752  &  0.33     & CAM-15                         \\
KOI-5 = TOI-1241     & G       &  5861  & 1.14   &  ...    &   ...     & BAT-13                         \\
LTT~1445             & M2.5~V  &  3337  & 0.257  &  0.276  &  0.0085   & WIN-19, WIN-22                 \\
Psi$^1$~Draconis     & F5~IV-V &  6212  & 1.19   &  1.50   &  3.0      & PAS-01, END-16                 \\
WASP-8               & G6~V    &  5600  & 1.09   &  0.98   &  0.79     & SOU-20                         \\
WASP-12              & F       &  6360  & 1.43   &  1.66   &  4.05     & COL-17                         \\
\enddata
\tablecomments{Stars with the luminosity class omitted are almost certainly main-sequence
stars.  See references for details, including information about data uncertainties.
Further references can be found in Sect.~3.1.1.  For most stars, multiple references
are available yielding similar results.
}
\label{table08}
\end{deluxetable}


\clearpage

\thispagestyle{empty} 
\begin{landscape}
\tabletypesize{\scriptsize}
\begin{deluxetable}{lllcccccl}
\tablenum{9}
\tablecaption{Triple Systems, Confirmed --- Planetary Data, Single Planets}
\tablewidth{0pt}
\tablehead{
System Name  & Planet Name & Planet Type & Mass          & Radius        & Semimajor Axis & Orbital Period & Eccentricity & Reference \\
...          & ...         & ...         & ($M_{\rm J}$) & ($R_{\rm J}$) & (au)           & (d)            & ...          & ...
}
\startdata
16~Cygni             & 16~Cygni~Bb          & Jupiter-type  &    2.38   & ...     &  1.693   &  799.5  &   0.689      &  PLA-13   \\
51~Eridani           & 51~Eridani~b         & Jupiter-type  &    2.6    &  1.11   & 11.1     &   28.1  &   0.53       &  ROS-20   \\
91~Aquarii           & 91~Aquarii~Ab        & Jupiter-type  & $>$3.20   & ...     &  0.70    &  181.4  &   0.027      &  MIT-13   \\
94~Ceti              & 94~Ceti~Ab           & Jupiter-type  &    1.86   & ...     &  1.427   &  535.7  &   0.30       &  PLA-13   \\
Epsilon~Indi         & Epsilon~Indi~Ab      & Jupiter-type  &    3.25   & ...     & 11.55    & 16510   &   0.26       &  FEN-19   \\
HAT-P-8              & HAT-P-8Ab            & Jupiter-type  &    1.275  &  1.32   &  0.04387 &  3.0760 &   0.0        &  MAN-13   \\
HAT-P-57             & HAT-P-57Ab           & Jupiter-type  & $<$1.85   &  1.41   &  0.04060 &  2.4653 &   0.0        &  HAR-15   \\
HD~126614            & HD~126614Ab          & Jupiter-type  & $>$0.38   & ...     &  2.35    &  1244   &   0.41       &  HOW-10   \\
HD~132563            & HD~132563Bb          & Jupiter-type  & $>$1.49   & ...     &  2.62    &  1544   &   0.22       &  DES-11   \\
HD~178911            & HD~178911Bb          & Jupiter-type  & $>$7.35   & ...     &  0.345   & 71.51   &   0.139      &  BUT-06   \\
HD~185269            & HD~185269Ab          & Jupiter-type  & $>$1.01   &  1.26   &  0.077   &  6.8378 &   0.229      &  LUH-19   \\
HD~196050            & HD~196050Ab          & Jupiter-type  & $>$2.90   & ...     &  2.54    &  1378   &   0.228      &  BUT-06   \\
HD~2638 / HD~2567    & HD~2638b             & Jupiter-type  & $>$0.48   & ...     &  0.044   &  3.4442 &   0.041      &  MOU-05   \\
HD~40979             & HD~40979Ab           & Jupiter-type  & $>$3.83   & ...     &  0.855   &  263.8  &   0.269      &  BUT-06   \\
HD~41004             & HD~41004Ab           & Jupiter-type  & $>$2.56   & ...     &  1.70    &  963    &   0.74       &  ZUC-04   \\
Kelt-4               & Kelt-4Ab             & Jupiter-type  &    0.878  &  1.71   &  0.04321 &  2.9896 &   0.03       &  EAS-16   \\
Kepler-13            & Kepler-13Ab          & Jupiter-type  & $\sim$6.5 &  1.41   &  0.03641 &  1.7636 &   0.0        &  SHP-14   \\
KOI-5 = TOI-1241     & KOI-5Ab              & Neptune-type  &    0.179  &  0.63   &  0.05961 & $\sim$5 &   0.0~~[set] &  HIR-17   \\
Psi$^1$~Draconis     & Psi$^1$~Draconis~Bb  & Jupiter-type  & $>$1.53   & ...     &  4.43    &  3117   &   0.40       &  END-16   \\
WASP-12              & WASP-12Ab            & Jupiter-type  &    1.47   &  1.90   &  0.02340 &  1.0914 &   0.0~~[set] &  COL-17   \\
\enddata
\tablecomments{
See references for details, including information about data uncertainties.
In some cases, alternate references are available yielding similar results.
}
\label{table09}
\end{deluxetable}
\end{landscape}


\clearpage

\thispagestyle{empty} 
\begin{landscape}
\tabletypesize{\scriptsize}
\begin{deluxetable}{lllrrccll}
\tablenum{10}
\tablecaption{Triple Systems, Confirmed --- Planetary Data, Multiple Planets}
\tablewidth{0pt}
\tablehead{
System Name  & Planet Name & Planet Type & Mass     & Radius   & Semimajor Axis & Orbital Period & Eccentricity & Reference \\
...          & ...         & ...         & (below)  & (below)  & (au)           & (d)            & ...          &
}
\startdata
Alpha~Centauri & Proxima~Centauri~b & Earth-type / super-Earth   &         1.63~$M_\oplus$  &   1.30~$R_\oplus$  &   0.04856  &   11.18    &   0.109     & ANG-16   \\
Alpha~Centauri & Proxima~Centauri~c & super-Earth / mini-Neptune &          7.0~$M_\oplus$  &        ...         &   1.489    &   1928     &   0.04      & DAM-20   \\
Alpha~Centauri & Proxima~Centauri~d & sub-Earth                  &      $>$0.26~$M_\oplus$  &   0.81~$R_\oplus$  &   0.02885  &    5.122   &   0.04      & FAR-22   \\
Gliese~667     & Gliese~667Cb       & super-Earth                &       $>$5.6~$M_\oplus$  &        ...         &   0.05043  &    7.1999  &   0.15      & ...$^a$  \\
Gliese~667     & Gliese~667Cc       & super-Earth                &       $>$4.1~$M_\oplus$  &        ...         &   0.12501  &   28.10    &   0.27      & ...$^a$  \\
HD~65216       & HD~65216Ab         & Jupiter-type               &     $>$1.295~$M_{\rm J}$ &        ...         &   1.301    &    577.6   &   0.27      & MAY-04   \\
HD~65216       & HD~65216Ac         & Jupiter-type               &      $>$2.03~$M_{\rm J}$ &        ...         &   5.75     &    5370    &   0.17      & WIT-19   \\
K2-290         & K2-290Ab           & mini-Neptune               &      $<$21.1~$M_\oplus$  &   3.06~$R_\oplus$  &   0.0923   &    9.212   &   0~~[set]  & HJO-19   \\
K2-290         & K2-290Ac           & Jupiter-type               &        0.774~$M_{\rm J}$ &   1.01~$R_{\rm J}$ &   0.305    &   48.367   &   0~~[set]  & HJO-19   \\
Kepler-444     & Kepler-444Ab       & sub-Earth                  &     ...                  &   0.41~$R_\oplus$  &   ...      &    3.600   &   ...       & MIL-17   \\
Kepler-444     & Kepler-444Ac       & sub-Earth                  &     ...                  &   0.52~$R_\oplus$  &   ...      &    4.546   &   ...       & MIL-17   \\
Kepler-444     & Kepler-444Ad       & sub-Earth                  &        0.036~$M_\oplus$  &   0.54~$R_\oplus$  &   0.0600   &    6.189   &   0.18      & MIL-17   \\
Kepler-444     & Kepler-444Ae       & sub-Earth                  &        0.034~$M_\oplus$  &   0.55~$R_\oplus$  &   0.0696   &    7.743   &   0.10      & MIL-17   \\
Kepler-444     & Kepler-444Af       & sub-Earth                  &     ...                  &   0.77~$R_\oplus$  &   ...      &    9.741   &   ...       & MIL-17   \\
LTT~1445       & LTT~1445Ab         & Earth-type / super-Earth   &         2.87~$M_\oplus$  &  1.305~$R_\oplus$  &   0.03813  &    5.3588  &   0.110     & WIN-22   \\
LTT~1445       & LTT~1445Ac         & Earth-type / super-Earth   &         1.54~$M_\oplus$  &   1.15~$R_\oplus$  &   0.02661  &    3.1239  &   0.223     & WIN-22   \\
WASP-8         & WASP-8Ab           & Jupiter-type               &        2.216~$M_{\rm J}$ &  1.165~$R_{\rm J}$ &   0.08170  &    8.1587  &   0.310     & SOU-20   \\
WASP-8         & WASP-8Ac           & Jupiter-type               &      $>$9.45~$M_{\rm J}$ &        ...         &   5.28     &   4323     &   0~~[set]  & KNU-14   \\
\enddata
\tablecomments{
$^a$ANG-12, ANG-13, FER-14, ROB-14.
See references for details, including information about data uncertainties.
In some cases, alternate references are available yielding similar results.
For planets, referred to as Earth-type / super-Earth, their final classification
needs to await improved mass measurements as well as the consensus of the scientific community
about the lower mass limit of super-Earth planets.
}
\label{table10}
\end{deluxetable}
\end{landscape}


\clearpage

\thispagestyle{empty}
\begin{deluxetable}{lccccccc}
\tablenum{11}
\tablecaption{Stellar Spectral Type Distribution, Triple Systems}
\tablewidth{0pt}
\tablehead{
Spectral Type & \multicolumn{5}{c}{Number of Stars}            & FMS & Expected Value (MS) \\
...           & Total & Total (MS) & Star~1 & Star~2 & Star~3  & ... & ...
}
\startdata
 A           &   3  &  3  &  2  &  1 &  0 &  0.6\%  &  0.4    \\
 F           &  10  &  9  &  8  &  0 &  1 &  3.0\%  &  1.9    \\
 G           &  14  & 13  &  8  &  1 &  4 &  7.6\%  &  4.9    \\
 K           &  14  & 13  &  5  &  6 &  2 & 12.1\%  &  7.7    \\
 M           &  27  & 27  &  1  & 13 & 13 & 76.5\%  & 48.9    \\
Substellar   &   3  & ... &  0  &  1 &  2 &  ...    &  ...    \\
\enddata
\tablecomments{
Total (MS) refers to the combined number of main-sequence stars, noting that stars with
no reported luminosity class are assumed to be main-sequence stars as well.  Total refers
to all stars considering that regarding the main component, there are also four evolved stars,
i.e., three subgiants and one giant, that are also included. Unconfirmed systems (status UC)
have been disregarded.   FMS indicates the observed frequency of main-sequence stars
in the solar neighborhood; see text for details.
}
\label{table11}
\end{deluxetable}


\clearpage

\thispagestyle{empty} 
\begin{deluxetable}{lccc}
\tablenum{12}
\tablecaption{Statistics of Stellar Spectral Types}
\tablewidth{0pt}
\tablehead{
Case   & Spectral Code &  Standard Dev. & Spectral Type
}
\startdata
Main-sequence stars (general)                                 &    5.3       &  ...  & $\sim$M3~V    \\
Binary S-Type planet host stars ($d < 500$~au)                &    3.7       &  0.8  & $\sim$G7~V    \\
Binary S-Type planet host stars ($d < 100$~au)                &    4.0       &  0.9  & $\sim$K0~V    \\
Triple star systems, planet host stars                        &    3.3       &  1.0  & $\sim$G3~V    \\
Triple star systems, planet host stars ($a_{\rm p} < 0.1$~au) &    3.4       &  1.3  & $\sim$G4~V    \\
Triple star systems, planet host stars ($a_{\rm p} > 0.1$~au) &    3.3       &  0.7  & $\sim$G3~V    \\
\enddata
\tablecomments{
Data for the binary systems are based on \cite{pila19}.
}
\label{table12}
\end{deluxetable}


\clearpage





\begin{thebibliography}{}

\bibitem[Akeson et al.(2021)]{akes21}
Akeson, R., Beichman, C., Kervella, P., Fomalont, E., \& Benedict, G. F. 2021,
\aj, 162, 14 (AKE-21)

\bibitem[Anglada-Escud{\'e} et al.(2012)]{angl12}
Anglada-Escud{\'e}, G., Arriagada, P., Vogt, S. S., et al. 2012,
\apjl, 751, L16 (ANG-12)

\bibitem[Anglada-Escud{\'e} et al.(2013)]{angl13}
Anglada-Escud{\'e}, G., Tuomi, M., Gerlach, E., et al. 2013,
\aap, 556, A126 (ANG-13)

\bibitem[Anglada-Escud{\'e} et al.(2016)]{angl16}
Anglada-Escud{\'e}, G., Amado, P. J., Barnes, J., et al. 2016,
\nat, 536, 437 (ANG-16)

\bibitem[Ayres(2014)]{ayre14}
Ayres, T. R. 2014, \aj, 147, 59

\bibitem[Bailer-Jones et al.(2018)]{bail18}
Bailer-Jones, C. A. L., Rybizki, J., Fouesneau, M., Mantelet, G., \&
Andrae, R. 2018, \aj, 156, 58

\bibitem[Baines et al. (2018)]{bain18}
Baines, E. K., Armstrong, J. T., Schmitt, H. R., et al. 2018, \aj, 155, 30 (BAI-18)

\bibitem[Barnes et al.(2016)]{barn16}
Barnes, R., Deitrick, R., Luger, R., et al. 2016; arXiv:1608.06919

\bibitem[Batalha et al.(2013)]{bata13}
Batalha, N. M., Rowe, J. F., Bryson, S. T., et al. 2013, \apjs, 204, 24 (BAT-13)

\bibitem[Bechter et al.(2014)]{bech14}
Bechter, E. B., Crepp, J. R., Ngo, H., et al. 2014, \apj, 788, 2 (BEC-14)

\bibitem[Benedict et al.(1999)]{bene99}
Benedict, G. F., McArthur, B., Chappell, D. W., et al. 1999, \aj, 118, 1086

\bibitem[Benedict \& McArthur(2020)]{bene20}
Benedict, G. F., \& McArthur, B. E. 2020, RNAAS, 4, 86

\bibitem[Bessel(1991)]{bess91}
Bessel, M. S. 1991, \aj, 101, 662 (BES-91)

\bibitem[Beuermann et al.(2012)]{beue12}
Beuermann, K., Dreizler, S.,  Hessman, F. V., \& Deller, J. 2012, \aap, 543, A138

\bibitem[Bi et al.(2020)]{bi20}
Bi, J., van der Marel, N., Dong, R., et al. 2020, \apjl, 895, L18

\bibitem[Bixel \& Apai(2017)]{bixe17}
Bixel, A., \& Apai, D. 2017, \apjl, 836, L31

\bibitem[Bohn et al.(2020)]{bohn20}
Bohn, A. J., Southworth, J., Ginski, C., et al. 2020, \aap, 635, A73 (BOH-20)

\bibitem[Bonfanti et al.(2015)]{bonf15}
Bonfanti, A., Ortolani, S., Piotto, G., \& Nascimbeni, V. 2015, \aap, 575, A18 (BON-15)

\bibitem[Bonfanti et al.(2016)]{bonf16}
Bonfanti, A., Ortolani, \& Nascimbeni, V. 2016, \aap, 585, A5 (BON-16)

\bibitem[Bowler \& Hillenbrand(2015)]{bowl15}
Bowler, B. P., \& Hillenbrand, L. A. 2015, \apjl, 811, L30

\bibitem[Boyajian et al.(2012)]{boya12}
Boyajian, T. S., von Braun, K., van Belle, G., et al. 2012, \apj, 757, 112 (BOY-12)

\bibitem[Boyajian et al.(2013)]{boya13}
Boyajian, T. S., von Braun, K., van Belle, G., et al. 2013, \apj, 771, 40 (BOY-13)

\bibitem[Boyle \& Cuntz(2021)]{boyl21}
Boyle, L., \& Cuntz, M. 2021, RNAAS, 5, 285

\bibitem[Brown et al.(2018)]{brow18}
Brown, A. G. A., Vallenari, A., Prusti, T., et al.
({\it Gaia} Collaboration) 2018, \aap, 616, A1

\bibitem[Busetti et al.(2018)]{buse18}
Busetti, F., Beust, H., \& Harley, C. 2018, \aap, 619, A91

\bibitem[Butler et al.(2006)]{butl06}
Butler, R. P., Wright, J. T., Marcy, G. W., et al. 2006, \apj, 646, 505 (BUT-06)

\bibitem[Campante et al.(2015)]{camp15}
Campante, T. L., Barclay, T., Swift, J. J., et al. 2015, \apj, 799, 170 (CAM-15)

\bibitem[Chabrier(2003)]{chab03}
Chabrier, G. 2003, \pasp, 115, 763

\bibitem[Chavero et al.(2019)]{chav19}
Chavero, C., de la Reza, R., Ghezzi, L., et al. 2019, \mnras, 487, 3162 (CHA-19)

\bibitem[Cheetham et al.(2018)]{chee18}
Cheetham, A., S{\'e}gransan, D., Peretti, S., et al. 2018, \aap, 614, A16 (CHE-18)

\bibitem[Ciardi(2021)]{ciar21}
Ciardi, D. 2021, \baas, 53, id. 239.02 (CIA-21)

\bibitem[Cochran et al.(1997)]{coch97}
Cochran, W. D., Hatzes, A. P., Butler, R. P., \& Marcy, G. W. 1997, \apj, 483, 457 (COC-97)

\bibitem[Collins et al.(2017)]{coll17}
Collins, K. A., Kielkopf, J. F., \& Stassun, K. G. 2017, \aj, 153, 78 (COL-17)

\bibitem[Colton et al.(2021)]{colt21}
Colton, N. M., Horch, E. P., Everett, M. E., et al. 2021, \aj, 161, 21

\bibitem[Correia et al.(2016)]{corr16}
Correia, A. C. M., Bou{\'e}, G., \& Laskar, J. 2016, CeMDA, 126, 189

\bibitem[Cuntz(2014)]{cunt14}
Cuntz, M. 2014, \apj, 780, 45

\bibitem[Cuntz \& Guinan(2016)]{cunt16}
Cuntz, M., \& Guinan, E. F. 2016, \apj, 827, 79

\bibitem[Cuntz \& Wang(2018)]{cunt18}
Cuntz, M., \& Wang, Zh. 2018, RNAAS, 2a, 19

\bibitem[Cuntz et al.(2007)]{cunt07}
Cuntz, M., Eberle, J., \& Musielak, Z. E. 2007, \apjl, 669, L105

\bibitem[Currie et al.(2012)]{curr12}
Currie, T., Debes, J., Rodigas, T. J., et al. 2012, \apjl, 760, L32

\bibitem[Czekala et al.(2019)]{czek19}
Czekala, I., Chiang, E., Andrews, S. M., et al. 2019, \apj, 883, 22

\bibitem[Damasso et al.(2020)]{dama20}
Damasso, M., Del Sordo, F., Anglada-Escud{\'e}, G., et al. 2020,
Science Adv., 6, 7467 (DAM-20)

\bibitem[Deacon et al.(2014)]{deac14}
Deacon, N. R., Liu, M. C., Magnier, E. A., et al. 2014, \apj, 792, 119 (DEA-14)

\bibitem[Delfosse et al.(2000)]{delf00}
Delfosse, X., S{\'e}gransan, D., Forveille, T., et al. 2000, \aap, 364, 217

\bibitem[Delfosse et al.(2013)]{delf13}
Delfosse, X., Bonfils, X., Forveille, T., et al. 2013, \aap, 553, A8

\bibitem[Demory et al.(2009)]{demo09}
Demory, B.-O., S{\'e}gransan, D., Forveille, T., et al. 2009, \aap, 505, 205 (DEM-09)

\bibitem[De Rosa et al.(2020)]{dero20}
De Rosa, R. J., Nielsen, E. L., Wang, J. J., et al. 2020, \aj, 159, 1 (ROS-20)

\bibitem[Desidera et al.(2004)]{desi04}
Desidera, S., Gratton, R. G., Scuderi, S., et al. 2004, \aap, 420, 683 (DES-04)

\bibitem[Desidera et al.(2011)]{desi11}
Desidera, S., Carolo, E., Gratton, R., et al. 2011, \aap, 533, A90 (DES-11)

\bibitem[Dieterich et al.(2018)]{diet18}
Dieterich, S. B., Weinberger, A. J., Boss, A. P., et al. 2018, \apj, 865, 28 (DIE-18)

\bibitem[Dumusque et al.(2012)]{dumu12}
Dumusque, X., Pepe, F., Lovis, C., et al. 2012, \nat, 491, 207

\bibitem[Dupuy et al.(2016)]{dupu16}
Dupuy, T. J., Kratter, K. M., Kraus, A. L., et al. 2016, \apj, 817, 80 (DUP-16)

\bibitem[Duquennoy \& Mayor(1991)]{duqu91}
Duquennoy, A., \& Mayor, M. 1991, \aap, 248, 485

\bibitem[Dutrey et al.(2014)]{dutr14}
Dutrey, A., di Folco, E., Guilloteau, S., et al. 2014, \nat, 514, 600

\bibitem[Dvorak(1982)]{dvor82}
Dvorak, R. 1982, OAWMN, 191, 423

\bibitem[Eastman et al.(2016)]{east16}
Eastman, J. D., Beatty, T. G., Siverd, R. J., et al. 2016, \aj, 151, 45 (EAS-16)

\bibitem[Eggenberger et al.(2007)]{egge07}
Eggenberger, A., Udry, S., Mazeh, T., Segal, Y., \& Mayor, M. 2007, \aap, 466, 1179

\bibitem[Eggleton \& Kiseleva(1995)]{eggl95}
Eggleton, P., \& Kiseleva, L. 1995, \apj, 455, 640

\bibitem[Eggleton \& Tokovinin(2008)]{eggl08}
Eggleton, P. P., \& Tokovinin, A. A. 2008, \mnras, 389, 869

\bibitem[Eggenberger et al.(2004)]{egge04}
Eggenberger, A., Udry, S., \& Mayor, M. 2004, \aap, 417, 353

\bibitem[Endl et al.(2016)]{endl16}
Endl, M., Brugamyer, E. J., Cochran, W. D., et al. 2016, \apj, 818, 34 (END-16)

\bibitem[Faria et al.(2022)]{fari22}
Faria, J. P., Su{\'a}rez Mascare{\~ n}o, A., Figueira, P., et al. 2022,
\aap, 658, A115 (FAR-22)

\bibitem[Fatuzzo et al.(2006)]{fatu06}
Fatuzzo, M., Adams, F. C., Gauvin, R., \& Proszkow, E. M. 2006, \pasp, 118, 849

\bibitem[Feigelson et al.(2006)]{feig06}
Feigelson, E. D., Lawson, W. A., Stark, M., Townsley, L., \& Garmire, G. P. 2006,
\aj, 131, 1730 (FEI-06)

\bibitem[Feng et al.(2019)]{feng19}
Feng, F., Anglada-Escud{\'e}, G., Tuomi, M., et al. 2019, \mnras, 490, 5002 (FEN-19)

\bibitem[Feroz \& Hobson(2014)]{fero14}
Feroz, F., \& Hobson, M. P. 2014, \mnras, 437, 3540 (FER-14)

\bibitem[Fischer et al.(2003)]{fisc03}
Fischer, D. A., Marcy, G. W., Butler, R. P., et al. 2003, \apj, 586, 1394 (FIS-03)

\bibitem[Ford et al.(2000)]{ford00}
Ford, E. B., Kozinsky, B., \& Rasio, F. A. 2000, \apj, 535, 385

\bibitem[Fuhrmann(2008)]{fuhr08}
Fuhrmann, K. 2008, \mnras, 384, 173 (FUH-08)

\bibitem[Fuhrmeister et al.(2011)]{fuhr11}
Fuhrmeister, B., Lalitha, S., Poppenhaeger, K., et al. 2011, \aap, 534, A133

\bibitem[G\'asp\'ar \& Rieke(2020)]{gasp20}
G\'asp\'ar, A., \& Rieke, G. H. 2020, PNAS, 117, 9712

\bibitem[Gaudi et al.(2020)]{gaud20}
Gaudi, B. S., Seager, S., Mennesson, B., et al. 2020, HabEx Report; arXiv:2001.06683

\bibitem[Ginski et al.(2016)]{gins16}
Ginski, C., Mugrauer, M., Seeliger, M., et al. 2016, \mnras, 457, 2173 (GIN-16)

\bibitem[Gray(2005)]{gray05}
Gray, D. F. 2005, The Observation and Analysis of Stellar Photosphere, 3rd edn., Cambridge Univ. Press, UK

\bibitem[Guenther et al.(2009)]{guen09}
Guenther, E. W., Hartmann, M., Esposito, M., et al. 2009, \aap, 507, 1659 (GUE-09)

\bibitem[Gullikson et al.(2015)]{gull15}
Gullikson, K., Endl, M., Cochran, W. D., \& MacQueen, P. J. 2015, \apj, 815, 62 (GUL-15)

\bibitem[Hamers et al.(2015)]{hame15}
Hamers, A. S., Perets, H. B., \& Portegies Zwart, S. F. 2015, \mnras, 455, 3180

\bibitem[Hartkopf et al.(2008)]{hart08}
Hartkopf, W. I., Mason, B. D., \& Rafferty, T. J. 2008, \aj, 135, 1334

\bibitem[Hartman et al.(2015)]{hart15}
Hartman, J. D., Bakos, G. {\'A}., Buchhave, L. A., et al. 2015, \aj, 150, 197 (HAR-15)

\bibitem[Hebb et al.(2009)]{hebb09}
Hebb, L., Collier Cameron, A., Loeillet, B., et al. 2009, \apj, 693, 1920 (HEB-09)

\bibitem[Hessman et al.(2010)]{hess10}
Hessman, F. V., Dhillon, V. S., Winget, D. E., et al. 2010; arXiv:1012.0707

\bibitem[Hirsch et al.(2017)]{hirs17}
Hirsch, L. A., Ciardi, D. R., Howard, A. W., et al. 2017, \aj, 153, 117

\bibitem[Hjorth et al.(2019)]{hjor19}
Hjorth, M., Justesen, A. B., Hirano, T., et al. 2019, \mnras, 484, 3522 (HJO-19)

\bibitem[Holman \& Wiegert(1999)]{holm99}
Holman, M. J., \& Wiegert, P. A. 1999, \aj, 117, 621

\bibitem[Horner et al.(2012)]{horn12}
Horner, J., Hinse, T. C., Wittenmyer, R. A., Marshall, J. P., \& Tinney, C. G.
2012, \mnras, 487, 2812

\bibitem[Howard et al.(2010)]{howa10}
Howard, A. W., Marcy, G. W., Johnson, J. A., et al. 2010, Science, 330, 653 (HOW-10)

\bibitem[Howard et al.(2018)]{howa18}
Howard, W. S., Tilley, M. A., Corbett, H., et al. 2018, \apjl, 860, L30

\bibitem[Howell et al.(2019)]{howe19}
Howell, S. B., Scott, N. J., Matson, R. A., Horch, E. P., \& Stephens, A.
2019, \aj, 158, 113

\bibitem[Jenson \& Akeson(2014)]{jens14}
Jenson, E. L. N., \& Akeson, R. 2014, \nat, 511, 567

\bibitem[Jones et al.(2002)]{jone02}
Jones, H. R. A., Butler, R. P., Marcy, G. W., et al. 2002, \mnras, 337, 1170 (JON-02)

\bibitem[Joyce \& Chaboyer(2018)]{joyc18}
Joyce, M., \& Chaboyer, B. 2018, \apj, 864, 99

\bibitem[Kalas et al.(2008)]{kala08}
Kalas, P., Graham, J. R., Chiang, E., et al. 2008, Science, 322, 1345

\bibitem[Kane et al.(2012)]{kane12}
Kane, S. R., Ciardi, D. R., Gelino, D. M., \& von Brown, K. 2012, \mnras, 425, 757

\bibitem[Kervella et al.(2017a)]{kerv17a}
Kervella, P., Bigot, L., Gallenne, A., \& Th{\'e}venin, F. 2017, \aap, 597, A137

\bibitem[Kervella et al.(2017b)]{kerv17b}
Kervella, P., Th{\'e}venin, F., \& Lovis, C. 2017, \aap, 598, L7 (KER-17)

\bibitem[Kervella et al.(2020)]{kerv20}
Kervella, P., Arenou, F., \& Schneider, J. 2020, \aap, 635, L14

\bibitem[Kiefer et al.(2021)]{kief21}
Kiefer, F., H{\'e}brard, G., Lecavelier des Etangs, A., et al. 2021, \aap, 645, 7

\bibitem[Knutson et al.(2014)]{knut14}
Knutson, H. A., Fulton, B. J., Montet, B. T., et al. 2014, \apj, 785,126 (KNU-14)

\bibitem[Konacki(2005)]{kona05}
Konacki, M. 2005, \nat, 436, 230 (KON-05)

\bibitem[Kovtyukh et al.(2003)]{kovt03}
Kovtyukh, V. V., Soubiran, C., Belik, S. I., \& Gorlova, N. I. 2003, \aap, 411, 559 (KOV-03)

\bibitem[Kroupa(2001)]{krou01}
Kroupa, P. 2001, \mnras, 322, 231

\bibitem[Kroupa(2002)]{krou02}
Kroupa, P. 2002, Science, 295, 82

\bibitem[Lada(2006)]{lada06}
Lada, C. J. 2006, \apjl, 640, L63

\bibitem[Lafarga et al.(2021)]{lafa21}
Lafarga, M., Ribas, I., Reiners, A., et al. 2021, \aap, 652, A28

\bibitem[Lagrange et al.(2017)]{lagr17}
Lagrange, A.-M., Keppler, M., Beust, H., et al. 2017, \aap, 608, L9

\bibitem[Latham et al.(2009)]{lath09}
Latham, D. W., Bakos, G. {\'A}., Torres, G., et al. 2009, \apj, 704, 1107 (LAT-09)

\bibitem[Lawler et al.(2015)]{lawl15}
Lawler, S. M., Greenstreet, S., \& Gladman, B. 2015, \apjl, 802, L20

\bibitem[Ledrew(2001)]{ledr01}
Ledrew, G. 2001, J. Roy. Astron. Soc. Canada, 95, 32

\bibitem[Lee et al.(2009)]{lee09}
Lee, J. W., Kim, S.-L., Kim, C.-H., et al. 2009, \aj, 137, 3181 (LEE-09)

\bibitem[Lehmann et al.(2016)]{lehm16}
Lehmann, H., Borkovits, T., Rappaport, S. A., et al. 2016, \apj, 819, 33 (LEH-16)

\bibitem[Lester et al.(2021)]{lest21}
Lester, K. V., Matson, R. A., Howell, S. B., et al. 2021, \aj, 162, 75

\bibitem[Lodieu et al.(2014)]{lodi14}
Lodieu, N., P{\'e}rez-Garrido, A., B{\'e}jar, V. J. S., et al. 2014, \aap, 569, A120
(LOD-14)

\bibitem[Luhn et al.(2019)]{luhn19}
Luhn, J. K., Bastien, F. A., Wright, J. T., et al. 2019, \aj, 157, 149 (LUH-19)

\bibitem[Ma et al.(2018)]{ma18}
Ma, B., Ge, J., Muterspaugh, M., et al. 2018, \mnras, 480, 2411 (MA-18)

\bibitem[Macintosh et al.(2015)]{maci15}
Macintosh, B., Graham, J. R., Barman, T., et al. 2015, Science, 350, 64
(MAC-15)

\bibitem[Mancini et al.(2013)]{manc13}
Mancini, L., Southworth, J., Ciceri, S., et al. 2013, \aap, 551, A11
(MAN-13)

\bibitem[Mann et al.(2015)]{mann15}
Mann, A. W., Feiden, G. A., Gaidos, E., Boyajian, T., \& von Braun, K. 2015,
\apj, 804, 64 (MAN-15)

\bibitem[Marcy et al.(2004)]{marc04}
Marcy, G. W., Butler, R. P., Fischer, D. A., \& Vogt, S. S. 2004, in
ASP Conf. Proc. 321, Extrasolar Planets: Today and Tomorrow, 
ed. J.-P. Beaulieu, et al. (San Francisco: ASP), 3

\bibitem[Mardling et al.(2001)]{mard01}
Mardling, R. A., \& Aarseth, S. J. 2001, \mnras, 321, 398

\bibitem[Marin \& Beluffi(2018)]{mari18}
Marin, F., \& Beluffi, C. 2018, J. British Interplanetary Soc., 71, 45

\bibitem[Mason et al.(2001)]{maso01}
Mason, B. D., Wycoff, G. L., Hartkopf, W. I., Douglass, G. G., \& Worley, C. E. 2001,
\aj, 122, 3466 (MAS-01)

\bibitem[Mason et al.(2017)]{maso17}
Mason, B. D., Hartkopf, W. I., \& Miles, K. N. 2017, \aj, 154, 200 (MAS-17)

\bibitem[Mayor \& Queloz(1995)]{mayo95}
Mayor, M., \& Queloz, D. 1995, \nat, 378, 355

\bibitem[Mayor et al.(2004)]{mayo04}
Mayor, M., Udry, S., Naef, D., et al. 2004, \aap, 415, 391 (MAY-04)

\bibitem[Meadows et al.(2018)]{mead18}
Meadows, V. S., Arney, G. N., Schwieterman, E. W., et al. 2018, Astrobiol., 18, 133

\bibitem[Metcalfe et al.(2015)]{metc15}
Metcalfe, T. S., Creevey, O. L., \& Davies, G. R. 2015, \apjl, 811, L37 (MET-15)

\bibitem[Mills \& Fabrycky(2017)]{mill17}
Mills, S. M., \& Fabrycky, D. C. 2017, \apjl, 838, L11 (MIL-17)

\bibitem[Mitchell et al.(2003)]{mitc03}
Mitchell, D. S., Frink, S., Quirrenbach, A., et al. 2003, \baas, 35, 1234 (MIT-03)

\bibitem[Mitchell et al.(2013)]{mitc13}
Mitchell, D. S., Reffert, S., Trifonov, T., Quirrenbach, A., \& Fischer, D. A.
2013, \aap, 555, A87 (MIT-13)

\bibitem[Montet et al.(2015)]{mont15}
Montet, B. T., Bowler, B. P., Shkolnik, E. L., et al. 2015, \apjl, 813, L11 (MON-15)

\bibitem[Morbey \& Brosterhus(1974)]{morb74}
Morbey, C. L., \& Brosterhus, E. B. 1974, \pasp, 86, 455

\bibitem[Moutou et al.(2005)]{mout05}
Moutou, C., Mayor, M., Bouchy, F., et al. 2005, \aap, 439, 367 (MOU-05)

\bibitem[Moutou et al.(2006)]{mout06}
Moutou, C., Loeillet, B., Bouchy, F., et al. 2006, \aap, 458, 327 (MOU-06)

\bibitem[Mugrauer(2019)]{mugr19}
Mugrauer, M. 2019, \mnras, 490, 5088

\bibitem[Mugrauer et al.(2005)]{mugr05}
Mugrauer, M., Neuh\"auser, R., Seifahrt, A., Mazeh, T., \& Guenther, E. 2005, \aap, 440, 1051
(MUG-05)

\bibitem[Mugrauer et al.(2007)]{mugr07}
Mugrauer, M., Neuh\"auser, R., \& Mazeh, T. 2007, \aap, 469, 755 (MUG-07)

\bibitem[Mugrauer et al.(2014)]{mugr14}
Mugrauer, M., Ginski, C., \& Seeliger, M. 2014, \mnras, 439, 1063

\bibitem[Myll\"ari et al.(2018)]{myll18}
Myll\"ari, A., Valtonen, M., Pasechnik, A., \& Mikkola, S. 2018, \mnras, 476, 830

\bibitem[Nielsen et al.(2017)]{niel17}
Nielsen, E. L., De Rosa, R. J., Rameau, J., et al. 2017, \aj, 154, 218

\bibitem[Pasinetti Fracassini et al.(2001)]{pasi01}
Pasinetti Fracassini, L. E., Pastori, L., Covino, S., \& Pozzi, A. 2001, \aap, 367, 521 (PAS-01)

\bibitem[Patience et al.(2002)]{pati02}
Patience, J., White, R. J., Ghez, A. M., et al. 2002, \apj, 581, 654

\bibitem[Pavlenko et al.(2017)]{pavl17}
Pavlenko, Y., Su{\'a}rez Mascare{\~ n}o, A., Rebolo, R., et al. 2017, \aap, 606, A49

\bibitem[Pilat-Lohinger et al.(2019)]{pila19}
Pilat-Lohinger, E., Eggl, S., \& Bazs{\'o}, {\'A}. 2019,
Planetary Habitability in Binary Systems, Advances in Planetary Science, Vol. 4
(Singapore: World Scientific)

\bibitem[Portegies Zwart \& McMillan(2005)]{port05}
Portegies Zwart, S. F., \& McMillan, S. L. W. 2005, \apjl, 633, L141 (POR-05)

\bibitem[Pourbaix \& Boffin(2016)]{pour16}
Pourbaix, D., \& Boffin, H. M. J. 2016, \aap, 586, A90

\bibitem[Quanz et al.(2021)]{quan21}
Quanz, S. P., Ottiger, M., Fontanet, E., et al. 2021, \aap, in press; arXiv:2101.07500

\bibitem[Queloz et al.(2010)]{quel10}
Queloz, D., Anderson, D. R., Collier Cameron, A., et al. 2010, \aap, 517, L1 (QUE-10)

\bibitem[Quintana et al.(2002)]{quin02}
Quintana, E. V., Lissauer, J. J., Chambers, J. E., \& Duncan, M. J. 2002, \apj, 576, 982

\bibitem[Raghavan et al.(2006)]{ragh06}
Raghavan, D., Henry, T. J., Mason, B. D., et al. 2006, \apj, 646, 523 (RAG-06)

\bibitem[Raghavan et al.(2010)]{ragh10}
Raghavan, D., McAlister, H. A., Henry, T. J., et al. 2010, \apjs, 190, 1

\bibitem[Rains et al.(2020)]{rain20}
Rains, A. D., Ireland, M. J., White, T. R., Casagrande, L., \& Karovicova, I.
2020, \mnras, 493, 2377 (RAI-20)

\bibitem[Rajan et al.(2017)]{raja17}
Rajan, A., Rameau, J., De Rosa, R. J., et al. 2017, \aj, 154, 10 (RAJ-17)

\bibitem[Rajpaul et al.(2016)]{rajp16}
Rajpaul, V., Aigrain, S., \& Roberts, S. 2016, \mnras, 456, L6

\bibitem[Ribas et al.(2017)]{riba17}
Ribas, I., Gregg, M. D., Boyajian, T. S., \& Bolmont, E. 2017, \aap, 603, A58 (RIB-17)

\bibitem[Roberts et al.(2011)]{robe11}
Roberts Jr., L. C., Turner, N. H., ten Brummelaar, T. A., Mason, B. D., \& Hartkopf, W. I.
2011, \aj, 142, 175 (ROB-11)

\bibitem[Roberts et al.(2015)]{robe15}
Roberts Jr., L. C., Tokovinin, A., Mason, B. D., et al. 2015, \aj, 149, 118 (ROB-15)

\bibitem[Robertson \& Mahadevan(2014)]{robe14}
Robertson, P., \& Mahadevan, S. 2014, \apjl, 793, L24 (ROB-14)

\bibitem[Robertson et al.(2016)]{robe16}
Robertson, P., Bender, C., Mahadevan, S., Roy, A., \& Ramsey, L. W. 2016, \apj, 832, 112

\bibitem[Rodr{\'{\i}}guez et al.(1998)]{rodr98}
Rodr{\'{\i}}guez, L. F., D'Alessio, P., Wilner, D. J., et al. 1998, \nat, 395, 355

\bibitem[Roell et al.(2012)]{roel12}
Roell, T., Neuh\"auser, R., Seifahrt, A., \& Mugrauer, M. 2012,
\aap, 542, A92 (ROE-12)

\bibitem[Rosenthal et al.(2021)]{ros21}
Rosenthal, L. J., Fulton, B. J., Hirsch, L. A., et al. 2021, \apjs, 255, 8

\bibitem[Santos et al.(2002)]{sant02}
Santos, N. C., Mayor, M., Naef, D., et al. 2002, \aap, 392, 215 (SAN-02)

\bibitem[Scholz et al.(2003)]{scho03}
Scholz, R.-D., McCaughrean, M. J., Lodieu, N., \& Kuhlbrodt, B. 2003,
\aap, 398, L29 (SCH-03)

\bibitem[Schwamb et al.(2013)]{schw13}
Schwamb, M. E., Orosz, J. A., Carter, J. A., et al. 2013, \apj, 768, 127 (SCH-13)

\bibitem[S{\'e}gransan et al.(2003)]{segr03}
S{\'e}gransan, D., Kervella, P., Forveille, T., \& Queloz, D. 2003, \aap, 397, L5 (SEG-03)

\bibitem[Shporer et al.(2011)]{shpo11}
Shporer, A., Jenkins, J. M., Rowe, J. F., et al. 2011, \aj, 142, 195 (SHP-11)

\bibitem[Shporer et al.(2014)]{shpo14}
Shporer, A., O'Rourke, J. G., Knutson, H. A., et al. 2014, \apj, 788, 92 (SHP-14)

\bibitem[Simon \& Schaefer(2011)]{simo11}
Simon, M., \& Schaefer, G. H. 2011, \apj, 743, 158 (SIM-11)

\bibitem[Smallwood et al.(2021)]{smal21}
Smallwood, J. L., Nealon, R., Chen, C., et al. 2021, \mnras, 508, 392

\bibitem[Sousa et al.(2018)]{sous18}
Sousa, S. G., Adibekyan, V., Delgado-Mena, E., et al. 2018, \aap, 620, A58 (SOU-18)

\bibitem[Southworth et al.(2020)]{sout20}
Southworth, J., Bohn, A. J., Kenworthy, M. A., Ginski, C., \& Mancini, L. 2020,
\aap, 635, A74 (SOU-20)

\bibitem[Stassun et al.(2017)]{stas17}
Stassun, K. G., Collins, K. A., \& Gaudi, B. S. 2017, \aj, 153, 136 (STA-17)

\bibitem[Stassun et al.(2019)]{stas19}
Stassun, K. G., Oelkers, R. J., Paegert, M., et al. 2019, \aj, 158, 138

\bibitem[Tamuz et al.(2008)]{tamu08}
Tamuz, O., S{\'e}gransan, D., Udry, S., et al. 2008, \aap, 480, L33 (TAM-08)

\bibitem[LUVOIR Team(2019)]{luvo19}
The LUVOIR Team 2019, The LUVOIR Mission Concept Study Final Report; arXiv:1912.06219

\bibitem[Th{\'e}venin et al.(2002)]{thev02}
Th{\'e}venin, F., Provost, J., Morel, P., et al. 2002, \aap, 392, L9

\bibitem[Todorov et al.(2010)]{todo10}
Todorov, K., Luhman, K. L., \& McLeod, K. K. 2010, \apjl, 714, L84 (TOD-10)

\bibitem[Tokovinin(2014a)]{toko14a}
Tokovinin, A. 2014a, \aj, 147, 86

\bibitem[Tokovinin(2014b)]{toko14b}
Tokovinin, A. 2014b, \aj, 147, 87

\bibitem[Tokovinin(1997)]{toko97}
Tokovinin, A. A. 1997, \aaps, 124, 75

\bibitem[Tokovinin(2008)]{toko08}
Tokovinin, A. 2008, \mnras, 389, 925

\bibitem[Torres et al.(2006)]{torr06}
Torres, C. A. O., Quast, G. R., da Silva, L., et al. 2006, \aap, 460, 695

\bibitem[Torres et al.(2012)]{torr12}
Torres, G., Fischer, D. A., Sozzetti, A., et al. 2012, \apj, 757, 161

\bibitem[Tucci Maia et al.(2014)]{tucc14}
Tucci Maia, M., Mel{\'e}ndez, J., \& Ram{\'{\i}}rez, I. 2014, \apjl, 790, L25 (TUC-14)

\bibitem[Tuomi et al.(2019)]{tuom19}
Tuomi, M., Jones, H. R. A., Butler, R. P., et al. 2019, \apjs, submitted; arXiv:1906.04644

\bibitem[van Belle \& von Braun(2009)]{bell09}
van Belle, G. T., \& von Braun, K. 2009, \apj, 694, 1085 (BEL-09)

\bibitem[Verrier \& Evans(2007)]{verr07}
Verrier, P. E., \& Evans, N. W. 2007, \mnras, 382, 1432

\bibitem[Wagner et al.(2016)]{wagn16}
Wagner, K., Apai, D., Kasper, M., et al. 2016, Science, 353, 673; retracted

\bibitem[Wagner et al.(2021)]{wagn21}
Wagner, K., Boehle, A., Pathak, P., et al. 2021, {\nat}~Comm. 12, 922

\bibitem[Wang et al.(2015a)]{wang15a}
Wang, J., Fischer, D. A., Horch, E. P., \& Huang, X.
2015a, \aj, 799, 229

\bibitem[Wang et al.(2015b)]{wang15b}
Wang, J., Fischer, D. A., Horch, E. P., \& Xie, J.-W.
2015b, \aj, 806, 248

\bibitem[Wang et al.(2014)]{wang14}
Wang, J., Fischer, D. A., Xie, J.-W., \& Ciardi, D. R. 2014, \apj, 791, 111

\bibitem[Wang et al.(2021)]{wang21}
Wang, X.-Y., Wang, Y.-H., Wang, S., et al. 2021, \apjs, 255, 15 (WAN-21)

\bibitem[Wang et al.(2022)]{wang22}
Wang, H. S., Lineweaver, C. H., Quanz, S. P., et al. 2022, \apj, 927, 134

\bibitem[White et al.(2013)]{whit13}
White, T. R., Huber, D., Maestro, V., et al. 2013, \mnras, 433, 1262

\bibitem[Winters et al.(2019)]{wint19}
Winters, J. G., Medina, A. A., Irwin, J. M., et al. 2019, \aj, 158, 152 (WIN-19)

\bibitem[Winters et al.(2022)]{wint22}
Winters, J. G., Cloutier, R., Medina, A. A., et al. 2022, \apj, 163, 168 (WIN-22)

\bibitem[Wittenmyer et al.(2019)]{witt19}
Wittenmyer, R. A., Clark, J. T., Zhao, J., et al. 2019, \mnras, 484, 5859 (WIT-19)

\bibitem[Zhao et al.(2018)]{zhao18}
Zhao, L., Fischer, D. A., Brewer, J., Giguere, M., \& Rojas-Ayala, B. 2018, \aj, 155, 24

\bibitem[Zirm(2007)]{zirm07}
Zirm, H. 2007, IAU Commision 26 (Double Stars) Information Circular, 161, 1 (ZIR-07)

\bibitem[Zucker et al.(2002)]{zuck02}
Zucker, S., Naef, D., Latham, D. W., et al. 2002, \apj, 568, 363 (ZUC-02)

\bibitem[Zucker et al.(2004)]{zuck04}
Zucker, S., Mazeh, T., Santos, N. C., Udry, S., \& Mayor, M. 2004, \aap, 426, 695 (ZUC-04)

\end{thebibliography}
\end{document}